\documentclass[usenatbib,useAMS]{mnras}
\pdfoutput=1

\usepackage{amsmath}
\usepackage{graphicx}
\usepackage{tabularx}
\usepackage{rotating}
\usepackage{subfig}

\usepackage{ctable}
\usepackage{amssymb}
\usepackage{multirow}

\usepackage{times}

\usepackage{hyperref}
\hypersetup{bookmarksdepth=subsubsection}

\renewcommand{\vec}[1]{\boldsymbol{#1}}
\newcommand{\kms}{\textrm{ km s}^{-1}}
\newcommand{\dd}[0]{\>\textrm{d}}
\newcommand{\df}[0]{\textrm{d}}
\newcommand{\avg}[1]{\langle #1 \rangle}

\begin{document}

\title[Turbulent mixing layers in protostellar outflows]{Turbulent mixing layers in supersonic protostellar outflows, with application to DG~Tauri}

\author[M.~C.~White et al.]
	{M.~C.~White\thanks{\href{mailto:marc.white@anu.edu.au}{marc.white@anu.edu.au}},
	 G.~V.~Bicknell,
	 R.~S.~Sutherland,
	 R.~Salmeron,
	 P.~J.~McGregor\thanks{This work would not have been possible without the key involvement of Professor Peter McGregor. His leadership in the construction of the Near-infrared Integral Field Spectrograph on the Gemini North telescope, his commissioning observations of DG~Tau, and his insightful contributions to the subsequent research, were all vital to the success of this project. Peter sadly passed away in March 2015. We will always remember him and are grateful for his many contributions to astronomy, including his mentoring of students, as well as his kindness and good humour. He will be sorely missed.}\\
	 Research School of Astronomy \& Astrophysics, The Australian National University, Canberra, ACT 2611, Australia}
\maketitle

\begin{abstract}

Turbulent entrainment processes may play an important role in the outflows from young stellar objects at all stages of their evolution. In particular, lateral entrainment of ambient material by high-velocity, well-collimated protostellar jets may be the cause of the multiple emission-line velocity components observed in the microjet-scale outflows driven by classical T Tauri stars. Intermediate-velocity outflow components may be emitted by a turbulent, shock-excited mixing layer along the boundaries of the jet. We present a formalism for describing such a mixing layer based on Reynolds decomposition of quantities measuring fundamental properties of the gas. In this model, the molecular wind from large disc radii provides a continual supply of material for entrainment. We calculate the total stress profile in the mixing layer, which allows us to estimate the dissipation of turbulent energy, and hence the luminosity of the layer. We utilize {\sc mappings} {\sc IV} shock models to determine the fraction of total emission that occurs in [Fe\ {\sc ii}] 1.644 $\umu$m line emission in order to facilitate comparison to previous observations of the young stellar object DG~Tauri. Our model accurately estimates the luminosity and changes in mass outflow rate of the intermediate-velocity component of the DG~Tau approaching outflow. Therefore, we propose that this component represents a turbulent mixing layer surrounding the well-collimated jet in this object. Finally, we compare and contrast our model to previous work in the field.

\end{abstract}


\section{Introduction}

Outflows are a near-universal component of young stellar objects (YSOs) throughout their evolution. They play a major role in star formation, and drive both the CO outflows seen in early-stage forming stars \citep[e.g.][]{B96,RB97} and the Herbig-Haro flows emanating from more mature protostars \citep[e.g.][]{RB01}. These outflows are thought to be launched either from the protostellar surface \citep[e.g.][]{FDC06}, magnetocentrifugally from magnetic reconnection points near the circumstellar disc truncation radius \citep[the X-wind and related models;][]{Se94,Re09} or from the disc surface at larger radii \citep[disc winds;][]{BP82,PN83}. In fact, more than one launch mechanism may be in operation \citep[e.g.][]{L03}.

The advent of the \emph{Hubble Space Telescope} and adaptive optics on ground-based telescopes has allowed the few hundred au of protostellar outflows closest to the protostar to be studied. The outflows associated with optically-revealed T Tauri stars take the form of well-collimated `microjets'. The study of these microjets is important because it is thought that they should not have interacted with the wider interstellar medium so close to the star \citep[although interactions may occur if there is a remnant protostellar envelope, e.g.][]{MCW13b}. If so, such observations may provide information on the outflow before it significantly interacts with the ambient medium.

The small-scale outflows from YSOs typically show an onion-like kinematic structure in optical and near-infrared (NIR) forbidden lines, with a well-collimated, high-velocity jet surrounded by a less-collimated, intermediate-velocity component \citep[e.g.][]{HMS97,Woe02,Pe03b,CBP08,R-Ge12,CoGe13}. The nature of the high-velocity jets in many sources has been studied extensively, including searches for signs of jet rotation\\ \citep[][hereafter Paper I]{Be00,Ce04,Ce07,MCW13a}, recollimation shocks \citep[Paper I;][]{Ge05,Ge08,GML09,Be11,Sce13} and studies of the propagation of shock-excited moving knots \citep[Paper I;][]{Be96,Ree02,Pe03b,A-Ae11}. The nature of the intermediate-velocity emitting material is still debated. Many authors attribute this emission to the presence of an intermediate-velocity disc-wind outflow component \citep[e.g.][]{Pe11}, which bridges the gap in launch radii between low-velocity ($\lesssim 50\kms$) molecular winds \citep[e.g.][]{Te04,Be08,A-Ae14} and high-velocity ($>200\kms$) jets. However, such an explanation does not provide a natural mechanism for the generation of forbidden optical- and NIR-line emission, which is attributed to shock excitation \citep[e.g.][]{Ne02}. It has yet to be explained how a steady-state intermediate-velocity disc wind would undergo shock excitation with relatively uniform intensity along the observable length of the feature.

It has been proposed that the intermediate-velocity forbidden-line emission components (IVCs) of small-scale protostellar outflows result from the lateral entrainment of ambient material, or a disc wind, by the high-velocity jet \citep[e.g.][]{Pe03b}. This suggestion is based on the observation that in some objects, the spatial width of the intermediate-velocity component increases with distance from the central star. Entrainment would cause the formation of a turbulent mixing layer between the supersonic jet and the material surrounding it, which would become shock-excited and emit in forbidden lines \citep[e.g.][]{Bie99}. Such a layer naturally grows in thickness with distance along the jet, reproducing the observations of low- to intermediate-velocity forbidden-line emission components in protostellar outflows \citep[e.g.][]{CR91,RCC95}.

The entrainment explanation has fallen out of favour recently for two reasons. First, jet simulations show that the jet pushes the ambient medium aside as it is launched \citep{L-CR10}, preventing ambient material from interacting with the sides of the jet. However, as suggested by \citet{Pe03b} and \citet{MCW13a}, the presence of a wide-angle molecular wind surrounding the central jet could supply a constant reservoir of material for entrainment by the jet. Secondly, hypersonic jets, such as protostellar microjets, should not form lateral entrainment layers if they are regarded as high Mach number, purely hydrodynamic flows. The formation of turbulent mixing layers is driven by the action of the Kelvin-Helmholtz (KH) instability at the jet-ambient material interface, and the growth rate of the KH instability decreases as the Mach number difference between the flows increases \citep{Ce94,T08}. However, protostellar jets are expected to exhibit strong toroidal magnetic fields \citep[e.g.][]{Z07}. The alignment of these fields with respect to the interface between the jet and the surrounding material, and their perpendicularity to the flow, may destabilise the interface to the KH instability \citep[Paper I;][]{MP82,RE83}. Therefore, entrainment remains an open possibility in protostellar jets. For a more detailed discussion of this argument, see Paper I, \S 4.2 therein.

Lateral entrainment in protostellar jets has been investigated analytically by \citet{CR91} and \citet{RCC95}. Their models provide useful predictions of mass entrainment rates and radiative luminosities associated with mixing layers. These models involve an `entrainment efficiency' parameter, which determines how effectively ambient material is drawn into the mixing layer by the KH instability. They constrained this parameter using the results of laboratory jet experiments, particularly those of \citet{BE72}. More recently, theoretical and experimental work has focussed on the role of compressibility in supersonic mixing layers \citep[e.g.][]{PR88,VSL96,SZD00}. Whilst these authors do not determine an explicit `entrainment efficiency', they demonstrate that compressibility effects play an important role in the evolution of turbulent mixing layers, leading to an asymptotically steady mixing layer growth rate with increasing Mach number difference between the two flows. This effect will also assist in facilitating entrainment in highly supersonic outflows. This relationship is shown by forming an explicit prescription for the turbulent stress within the mixing layer. In this paper, we choose to develop an alternative semi-empirical approach to radiative mixing layers, which relies solely upon directly observable quantities. In this way, we generate estimates for mixing layer bulk properties from our model based on the observed parameters of YSO outflow components, and compare them to observation.

This paper is organized as follows. In \S\ref{sec:model-turb} we describe our model, which parametrizes the physical properties of the mixing layer using the observed layer growth rate. In \S\ref{sec:compare}, we first compute a grid of shock models to determine the ratio between the observable [Fe\ {\sc ii}] line emission of protostellar jet mixing layers, and the mixing layer bolometric luminosity estimated by our model. We then directly compare our model to the [Fe\ {\sc ii}] IVC of the approaching outflow from the YSO DG~Tauri, and find that it is in excellent agreement with observations. \S\ref{sec:disc} compares our model to the previous work of \citet{CR91} and \citet{RCC95}, and includes an estimation of the laminar jet length in DG~Tau. We summarize our work in \S\ref{sec:concl2}.

\section{Model}\label{sec:model-turb}

We construct an analytical, semi-empirical model of a two-dimensional turbulent entrainment layer in order to interpret the [Fe\ {\sc ii}] 1.644 $\umu$m IVC line emission observed in DG~Tau.\footnote{A two-dimensional model is an adequate representation of the jet edge, given that the mixing layer width is not significantly greater than the jet radius (see below).} The model describes the turbulent mixing layer that forms between a high-velocity jet and a low-velocity wider-angle wind, and depends only upon directly observable quantities, removing the requirement to specify an `entrainment efficiency' parameter \citep[e.g.][]{CR91,RCC95}. We use the observed spreading rate of the layer to calculate the dissipation of turbulent energy in the entrainment layer, and its resulting luminosity. The first step in this process is the calculation of total turbulent stress, $t_{xy}$, in the mixing layer.

\begin{figure}
\includegraphics[width=\columnwidth]{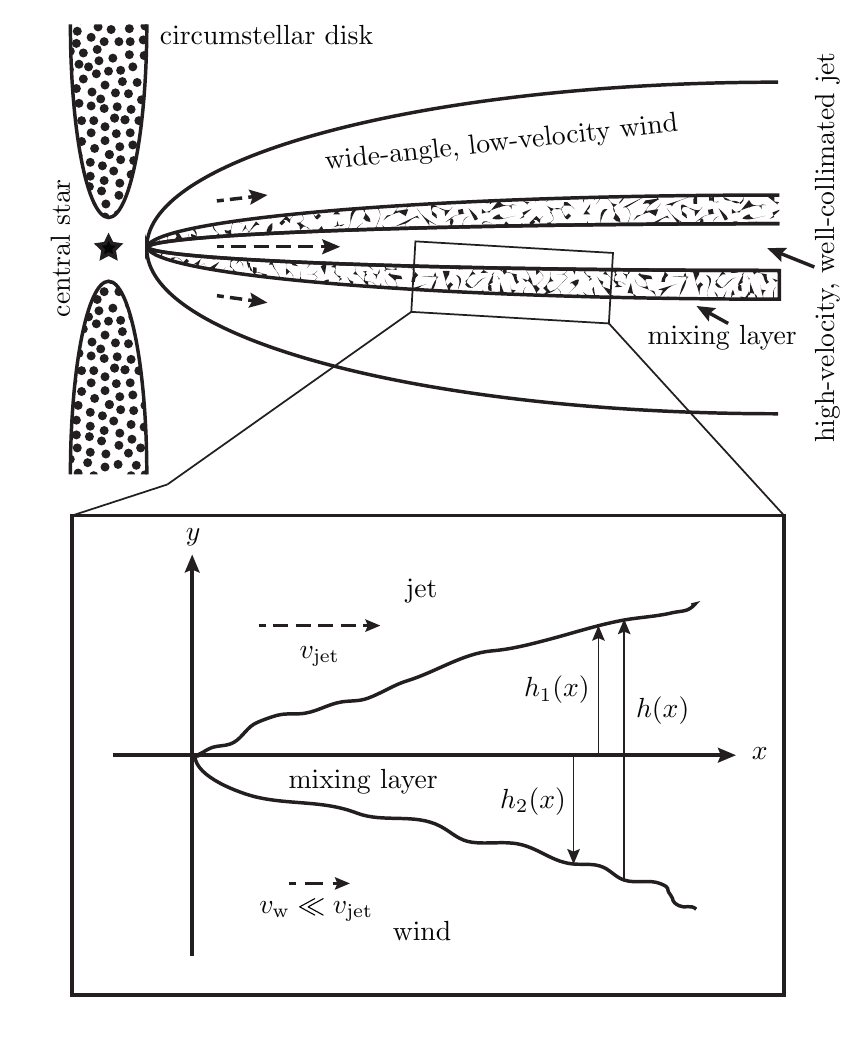}
\caption[Schematic of mixing layer model]{A representation of the model setup used throughout this paper. A high-velocity, well-collimated jet is launched from a protostar-circumstellar disc system. This jet is surrounded by a wider-angle disc wind (top panel). The interface between the jet and the wind is approximated by a two-dimensional turbulent shear layer (bottom panel). The $x$-axis of the model is placed where the jet-wind interface would lie in the absence of the mixing layer, and is parallel to the direction of the jet. Dashed arrows show the flow direction of the components. Model components are not to scale.}\label{fig:mixlayerdiag}
\end{figure}

The model setup is shown in Fig.~\ref{fig:mixlayerdiag}. A high-velocity jet with density $\rho_\textrm{jet}$ propagates at velocity $v_\textrm{jet}$ away from the star-disc system. The jet is surrounded by a wider-angle molecular wind, with density $\rho_\textrm{w}$ and velocity $v_\textrm{w}\ll v_\textrm{jet}$. We henceforth refer to this wind as the `ambient wind'. A turbulent mixing layer forms at the interface between the two flows, as a result of the KH instability. We approximate this interface with a two-dimensional model. In this model, the $x$-axis is defined as the unperturbed jet-wind boundary. This is the streamwise direction; the transverse coordinate is $y$. The mixing layer width, $h(x)$, increases monotonically with distance from the central star. We define the depth the mixing layer expands into the jet as $h_1(x)$, and the depth it penetrates the ambient wind as $h_2(x)$, where $h_2(x)<0$. Hence, $h(x)=h_1(x)-h_2(x)$.

An averaging prescription is used to describe the mean flow. We adopt the mass-weighted statistical averaging prescription of \citet{F69}, which was introduced to the study of astrophysical flows by \citet{Bi84}; see also \citet{KB04}. All time-varying quantities are decomposed into an average component and a fluctuating component. Quantities such as pressure, $p$, density, $\rho$, and magnetic field, $B$, are expressed in terms of mean (bar) and fluctuating (primed) components, such that the time-average of the fluctuating component (angle brackets) is zero:
\begin{alignat}{2}
p & = \bar{p}+p^\prime & \textrm{ where }  \avg{p^\prime} & =0\textrm{;} \\
\rho & = \bar{\rho}+\rho^\prime &\textrm{ where } \avg{\rho^\prime} & =0\textrm{; and} \\
B_i & = \bar{B}_i+B_i^\prime &\textrm{ where } \avg{B_i^\prime} & =0\textrm{,}
\end{alignat}
where subscripts $i$ and $j$ represent generalized coordinates. As prescribed by \citet{F69}, the velocity, $v_i$, is mass-weighted, and is expressed as
\begin{align}\label{eq:velavg}
v_i = \tilde{v}_i+v_i^\prime &\textrm{, where } \avg{\rho v_i^\prime} = 0\textrm{.}
\end{align}
This approach has two advantages. First, mass is conserved in the mean flow \citep{F69}. Secondly, it prevents the generation of an excessive number of terms when the dynamical equations are statistically averaged; e.g.~the mean value of the momentum flux is simply expressed as
\begin{equation}\label{eq:avgexample}
\avg{\rho v_iv_j}=\bar{\rho}\tilde{v}_i\tilde{v_j}+\avg{\rho v_i^\prime v_j^\prime}\textrm{.}
\end{equation}
This approach is common in fluid dynamics, and has been used in the theory of compressible turbulent jets and accretion discs \citep[see][]{Bi84,KB04}.

\subsection{Characteristic Equations}\label{sec:chareqn}

Consider a compressible magnetized fluid with density $\rho$, velocity $v$, pressure $p$, magnetic field $\vec{B}$, in a gravitational potential field $\phi_\textrm{G}$. Averaging the mass continuity and momentum conservation equations of magnetohydrodynamics (\S\ref{app:suppl-chareqn}) yields, for a quasi-steady state system,
\begin{align}
\frac{\partial (\bar{\rho}\tilde{v}_x)}{\partial x}+\frac{\partial (\bar{\rho}\tilde{v}_y)}{\partial y} & =0\textrm{, and} \label{eq:eocmixing}\\
\frac{\partial(\bar{\rho}\tilde{v}_i\tilde{v}_j)}{\partial x_j} & = -\bar{\rho}\frac{\partial \phi_\textrm{G}}{\partial x_i}-\frac{\partial\bar{p}}{\partial x_i}+\frac{\partial (t_{ij}^R+t_{ij}^B)}{\partial x_j}\textrm{.} \label{eq:commixing}
\end{align}
The magnetic stress tensor is defined as
\begin{equation}
t_{ij}^B=\frac{\avg{B_i^\prime B_j^\prime}}{4\pi}-\delta_{ij}\frac{\avg{B^{\prime 2}}}{8\pi}\textrm{,}
\end{equation}
assuming that the magnetic field is dominated by its turbulent component, so that $\bar{B}_i=0$. We define the Reynolds stress tensor as
\begin{equation}
t_{ij}^\textrm{R}=-\langle\rho v_i^\prime v_j^\prime \rangle
\end{equation}
\citep{KB04}. The total turbulent stress is
\begin{equation}\label{eq:totstress}
t_{ij}=t_{ij}^\textrm{R}+t_{ij}^B\textrm{.}
\end{equation}

The aim of our calculation is to estimate the mass entrainment rate and bolometric luminosity of the mixing layer, based on directly observed parameters. We consider the $i=x$ momentum equation, that is, the equation governing the streamwise evolution of momentum resulting from the lateral transfer of momentum within the mixing layer. We neglect the streamwise pressure, magnetic and gravitational gradients, so that
\begin{align}\label{eq:eommixing}
\frac{\partial (\bar{\rho}\tilde{v}_x^2)}{\partial x}+\frac{\partial(\bar{\rho}\tilde{v}_x\tilde{v}_y)}{\partial y} & \equiv \bar{\rho}\tilde{v}_x\frac{\partial \tilde{v}_x}{\partial x}+\bar{\rho}\tilde{v}_y\frac{\partial \tilde{v}_x}{\partial y} = \frac{\partial t_{xy}}{\partial y}\textrm{,}
\end{align}
where $t_{xy}=t_{xy}^R+t_{xy}^B$ as per equation (\ref{eq:totstress}). 

We analyse the orders of magnitude of the terms in equations (\ref{eq:eocmixing}) and (\ref{eq:eommixing}) in Appendix \ref{app:oom}. We denote the characteristic advective length scale of the mixing layer in the $x$-direction as $L$. In this analysis, we show that the following order of magnitude relationships exist in our model:
\begin{align}
\tilde{v}_y & \sim \frac{h}{L}\tilde{v}_x\textrm{,} \\
v^\prime & \sim \left(\frac{h}{L}\right)^{1/2} \tilde{v}_x\textrm{, and} \\
h^\prime & \sim \frac{h}{L}\textrm{.}
\end{align}
This analysis also justifies the neglecting of the Reynolds stress term governing the streamwise transfer of momentum, $\partial t_{xx}/\partial x$, in equation (\ref{eq:eommixing}).

In order to achieve our goal of estimating the rate of dissipation of turbulence, we define a pseudo self-similar variable in the transverse direction,
\begin{align}
\xi(x,y) & = \frac{y}{h(x)} \label{eq:zeta}\\
\Rightarrow \xi_1 & = \frac{y_1(x)}{h(x)}\textrm{, and }\xi_2(x) = \frac{y_2(x)}{h(x)}\textrm{.}
\end{align}
We note that $\xi_1-\xi_2=1$. It may also be shown that total (thermal plus turbulent) pressure balance is maintained across the mixing layer (\S\ref{app:oom}):
\begin{equation}
p_\textrm{tot} = \underbrace{\bar{p}}_{p_\textrm{thermal}} + \underbrace{\langle \rho v_y^{\prime 2} \rangle}_{p_\textrm{turbulent}} = \textrm{const.}
\end{equation}
This allows us to specify the quantity $\mathcal{W}$ (\S\ref{app:oom}, equation \ref{eq:W}), and streamwise velocity, $\tilde{v}_x$, in the mixing layer as a function of transverse position. As the simplest approximation, we prescribe linear profiles across the layer:
\begin{align}
\tilde{v}_x & = U(\xi) v_\textrm{jet}\textrm{, and} \label{eq:vprofile}\\
\mathcal{W} & = S(\xi)(\mathcal{W}_\textrm{jet} - \mathcal{W}_\textrm{w}) + \mathcal{W}_\textrm{w}\textrm{, where} \label{eq:Tprofile}\\
U(\xi) = S(\xi) & = \xi - \xi_2 = \xi-\xi_1+1{.}
\end{align}

\subsection{Transverse Density, Velocity and Turbulent Stress Profiles}\label{sec:model-turb-profiles}

We assume pressure equilibrium across the mixing layer.\footnote{In the case of DG~Tau, the presence of a stationary recollimation shock in the jet channel (Paper I) indicates that the jet is in pressure equilibrium with its environs downstream of this shock.} Let the jet-to-ambient wind density ratio be $\eta = \rho_\textrm{jet}/\rho_\textrm{w}$. The density is given by
\begin{equation}\label{eq:rhoprelation}
\bar{\rho} =\frac{\mu m}{k\mathcal{W}}\bar{p}_\textrm{tot}\textrm{,}
\end{equation}
where $k$ is the Boltzmann constant, $m$ is the atomic mass unit, and $\mu$ is the molecular weight of the gas. It follows that the mixing layer density profile is given by
\begin{equation}\label{eq:rhofrometa}
\bar{\rho}(\eta, \xi) =\frac{\rho_\textrm{jet}}{\eta + (1-\eta)S(\xi)}\textrm{.}
\end{equation}
It is important to note that the density and temperature profiles are intrinsically linked due to our assumption of constant pressure; it is not possible for one profile to vary across the mixing layer unless the other profile also varies. 

We now calculate the transverse velocity and turbulent stress profiles within the mixing layer. We transform equations (\ref{eq:eocmixing}) and (\ref{eq:eommixing}) into the $(x,\xi)$ coordinate system. The equation of continuity, equation (\ref{eq:eocmixing}), becomes
\begin{equation}\label{eq:eocxxi}
\frac{\partial (\bar{\rho}\tilde{v}_y)}{\partial\xi}=h^\prime(x)\xi\frac{\partial \bar{\rho}\tilde{v}_x}{\partial\xi}\textrm{.}
\end{equation}
We can calculate the tranverse velocity profile from this equation, after substituting in the density (equation \ref{eq:rhofrometa}) and streamwise velocity (equation \ref{eq:vprofile}) profiles. This gives:
\begin{align}\label{eq:vy}
\tilde{v}_y(\eta, x, \xi) &=v_\textrm{jet}h^\prime(x)\left[\eta +(1-\eta)S(\xi)\right] \nonumber\\
            &\phantom{=}\times\int_{\xi_1}^\xi\xi^\prime\frac{\df}{\df\xi^\prime}\left[\frac{U(\xi^\prime)}{\eta +(1-\eta)S(\xi^\prime)}\right]\dd\xi^\prime\textrm{.}
\end{align}
The integral factor in equation (\ref{eq:vy}) occurs in many of the subsequent expressions, and we define it as $\mathcal{D}(\eta,\xi)$. This factor has a closed-form solution (\S\ref{app:suppl-dimless}) after imposing the boundary condition $\tilde{v}_y(\xi_1)=0$, so the transverse velocity varies smoothly from $v_y=0$ in the jet into the mixing layer. This is a reasonable boundary condition, because the supersonic jet will approach the mixing layer boundary so quickly that it will not be substantially deflected by turbulence prior to impacting the mixing layer.

\begin{figure}
\includegraphics[width=\columnwidth]{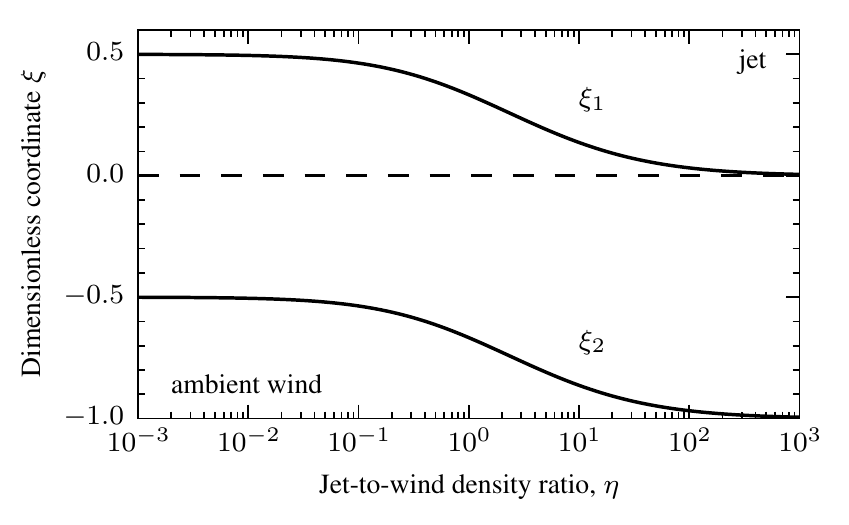}
\caption[Mixing layer boundary positions]{Position of the mixing layer boundaries with the jet ($\xi_1$) and with the ambient wind ($\xi_2$) as a function of the jet-to-wind density ratio, $\eta$. The position of the boundary between the jet and wind in the absence of a mixing layer is shown by the dashed line.}\label{fig:mixbounds}
\end{figure}

Following transformation into the $(x,\xi)$ coordinate system, the equation of downstream momentum conservation, equation (\ref{eq:eommixing}), can be rearranged to provide an equation for the turbulent stress:
\begin{equation}\label{eq:dxitij}
\frac{\partial t_{xy}}{\partial \xi}=\rho_\textrm{jet} v_j^2 h^\prime(x)\left(\frac{-(\xi-\xi_1+1)\xi}{\eta+(1-\eta)(\xi-\xi_1+1)}+\mathcal{D}(\eta,\xi)\right)\textrm{.}
\end{equation}
Integration of equation (\ref{eq:dxitij}) gives:
\begin{equation}\label{eq:txy}
t_{xy}(\eta, x, \xi) = \rho_\textrm{jet} v_\textrm{jet}^2 h^\prime(x)\mathcal{F}(\eta,\xi)\textrm{,}
\end{equation}
where the function $\mathcal{F}(\eta,\xi)$ is given in \S\ref{app:suppl-dimless}. We set $t_{xy}(\xi_1)=t_{xy}(\xi_2)=0$ since we expect the turbulence to be confined primarily to the region $\xi_2<\xi<\xi_1$. The condition $t_{xy}(\xi_1)=0$ is used to compute the form of $\mathcal{F}(\eta,\xi)$ (\S\ref{app:suppl-dimless}); the condition $t_{xy}(\xi_2)=0$, and hence $\mathcal{F}(\xi_2,\eta)=0$, allows us to calculate the position of the jet-mixing layer boundary, $\xi_1$, in $(x,\xi)$-space as a function of only the jet-to-wind density ratio $\eta$:
\begin{alignat}{2}
\xi_1(\eta) & =\frac{2\eta^2\log(\eta)+(4-3\eta)\eta-1}{2(\eta-1)^3} & \textrm{ for }\eta & \neq 1\textrm{, and}\nonumber\\
 &= \frac{1}{3} & \textrm{ for }\eta &=1\textrm{.} \label{eq:xi1eta}
\end{alignat}
The position of the mixing layer boundaries as a function of $\eta$ is shown in Fig.~\ref{fig:mixbounds}. In the limit of a significantly underdense jet ($\eta\rightarrow 0$), the mixing layer penetrates the jet and wind evenly. In the limit of a significantly overdense jet ($\eta\rightarrow\infty$), the mixing layer almost exclusively penetrates the ambient wind.

\begin{figure}
\includegraphics[width=\columnwidth]{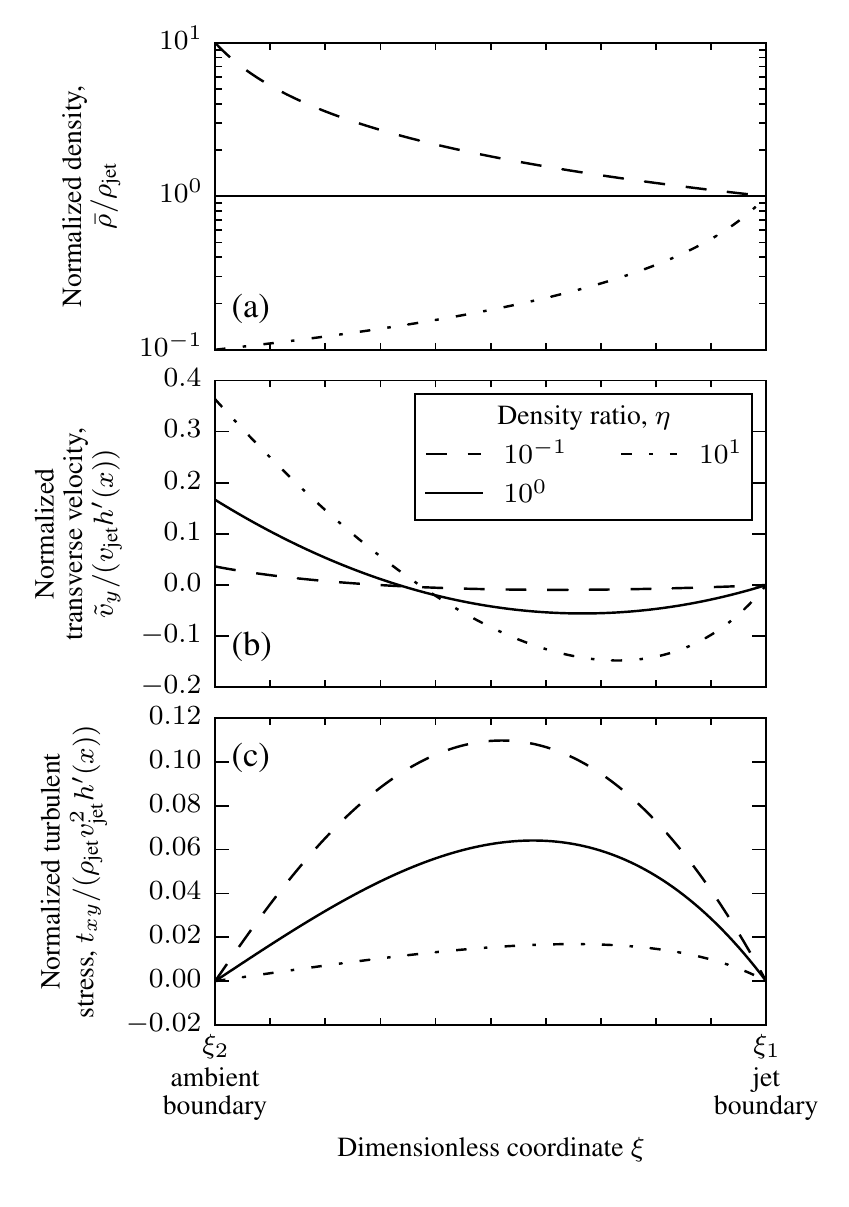}
\caption[Mixing layer density, transverse velocity and turbulent stress profiles]{Normalized (a) density, (b) transverse velocity and (c) turbulent stress profiles for the mixing layer described by this model. The profiles are plotted between the mixing layer boundaries $\xi_2$ and $\xi_1$. The numerical values that these boundaries take are different for different jet-to-wind density ratios $\eta$, as per equation (\ref{eq:xi1eta}) and Fig.~\ref{fig:mixbounds}.}\label{fig:mixprofiles}
\end{figure}

Knowledge of the position of the mixing layer boundaries allows the forms of the transverse velocity and turbulent stress profiles to be directly calculated as a function of position within the mixing layer, $\xi$, and the jet-to-wind density ratio, $\eta$. The forms of these expressions are algebraically complex, and are given in \S\ref{app:suppl-profiles}. The mixing layer density, transverse velocity and turbulent stress profiles are shown in Fig.~\ref{fig:mixprofiles}. 

\begin{figure}
\includegraphics[width=\columnwidth]{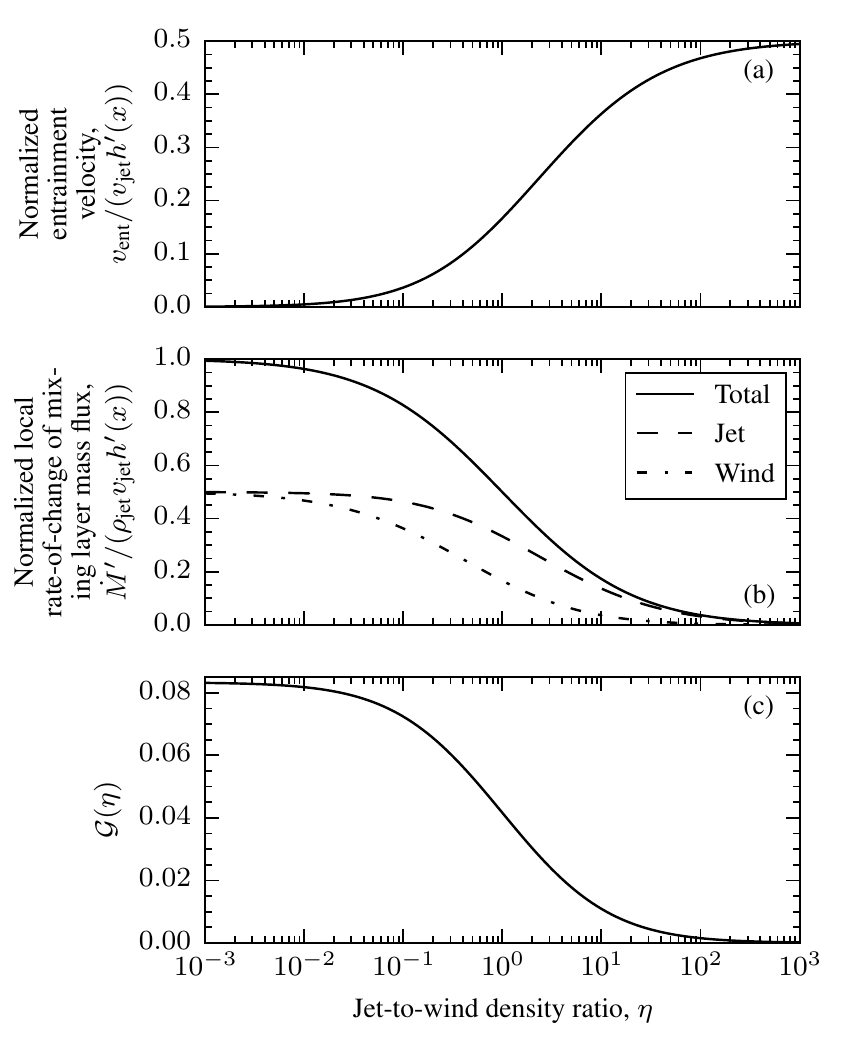}
\caption[Mixing layer entrainment velocity and mass gain]{Mixing layer parameters which only depend on the jet-to-ambient-wind density ratio, $\eta$. (a) Normalized entrainment velocity from the ambient wind, from equation (\ref{eq:vent}). (b) Normalized rate-of-change of the mixing layer mass flux, $\dot{M}^\prime$. The contribution to the rate-of-change of the mixing layer mass flux from jet interception and wind entrainment ($\dot{M}_\textrm{ent}^\prime$) is shown by the dashed and dot-dashed curves, respectively. (c) Dimensionless function $\mathcal{G}(\eta)$, for the determination of the rate of turbulent energy production per unit area, from equation (\ref{eq:Geta}).}\label{fig:MentG}
\end{figure}

As seen in Fig.~\ref{fig:mixprofiles}(b), ambient wind material is pulled upwards into the mixing layer at the wind-mixing layer boundary ($\xi_2$) with an effective entrainment velocity,
\begin{alignat}{2}
v_\textrm{ent} & = \tilde{v}_y(\eta, x, \xi_2) && \\
               & = v_\textrm{jet}h^\prime(x)\frac{\eta(\eta^2-2\eta\log(\eta)-1)}{2(\eta-1)^3} & \textrm{ for }\eta & \neq 1\textrm{, and}\nonumber\\
               & = v_\textrm{jet}h^\prime(x)\frac{1}{6} & \textrm{ for }\eta & =1\textrm{.}\label{eq:vent}
\end{alignat}
This is equivalent to the entrainment velocity specified in the models of \citet{CR91} and \citet{RCC95}. However, it occurs naturally as a result of the boundary conditions of the problem, rather than being specified by an experimentally-determined `entrainment efficiency' parameter. We compare the entrainment velocity of our model to the earlier work in \S\S\ref{sec:compare-vent} and \ref{sec:disc-prevmodels}. The normalized entrainment velocity, $v_y/(v_\textrm{jet}h^\prime(x))$, is plotted as a function of $\eta$ in Fig.~\ref{fig:MentG}(a).

\subsection{Mass Flux and Entrainment Rate}

We define the mixing layer mass flux to be
\begin{equation}\label{eq:Mdot}
\dot{M}(x)\equiv \int_{\xi_2}^{\xi_1}\rho(\xi)v_\textrm{jet}U(\xi)h(x)\dd\xi\textrm{.}
\end{equation}
The contribution to the mass flux from intercepted jet material is simply given by
\begin{equation}
\dot{M}_\textrm{jet}(x) =\rho_\textrm{jet}v_\textrm{jet}h(x)\xi_1(\eta)\textrm{.}
\end{equation}
\citep[cf.][]{RCC95}. Therefore, the entrained mass flux is
\begin{equation}\label{eq:Mdotent}
\dot{M}_\textrm{ent}(x)\equiv \int_{\xi_2}^{\xi_1}\rho(\xi)v_\textrm{jet}U(\xi)h(x)\dd\xi - \rho_\textrm{jet}v_\textrm{jet}h_1(x)\xi_1(\eta)\textrm{.}
\end{equation}
The mass entrainment rate from the ambient wind is simply the derivative of equation (\ref{eq:Mdotent}) with respect to $x$. It can be shown (\S\ref{app:suppl-Mentrate}) that 
\begin{equation}
\frac{\partial\dot{M}_\textrm{ent}}{\partial x}\equiv \dot{M}_\textrm{ent}^\prime=\rho_\textrm{w}v_\textrm{ent}\textrm{,}
\end{equation}
as expected, since ambient wind material is being drawn into the mixing layer with velocity $v_\textrm{ent}$ (\S\ref{sec:model-turb-profiles}).

\begin{figure}
\includegraphics[width=\columnwidth]{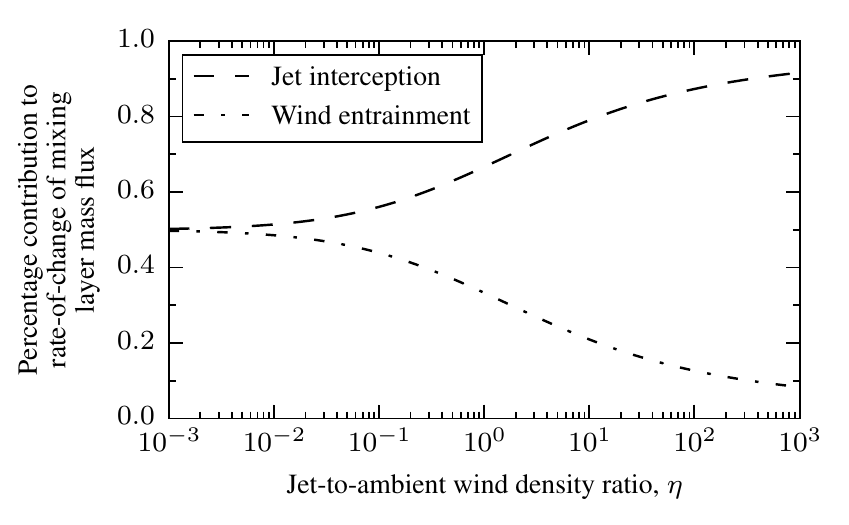}
\caption[Contribution to rate-of-change of mixing layer mass flux]{Percentage contribution to the rate-of-change of mixing layer mass flux from jet interception (dashed curve) and ambient wind entrainment ($\dot{M}_\textrm{ent}^\prime$, dot-dashed curve) as a function of jet-to-ambient-wind density ratio $\eta$.}\label{fig:Mentcontrib}
\end{figure}

The rate-of-change of the mixing layer mass flux, $\partial\dot{M}/\partial x\equiv\dot{M}^\prime$, is shown in Fig.~\ref{fig:MentG}(b). The contribution of wind entrainment to the mass flux of the mixing layer is greatest for an underdense jet (Fig.~\ref{fig:Mentcontrib}). 

\subsection{Turbulent Energy Production}\label{sec:model-turb-turbE}

The ultimate aim of this model is to determine the rate of turbulent energy production and subsequent dissipation and radiation in the mixing layer. The rate of turbulent energy production, $\dot{E}_\textrm{turb}=2 t_{xy}s_{xy}$, where $s_{xy}$ is the shear in the mixing layer \citep{KB04}. The mean shear may be calculated from
\begin{equation}
s_{xy} = \frac{1}{2}\left(\frac{\partial\tilde{v}_x}{\partial y}+\frac{\partial\tilde{v}_y}{\partial x}\right)\approx \frac{1}{2}\frac{\partial\tilde{v}_x}{\partial y}
\end{equation}
since the average transverse velocity, $\tilde{v}_y$, varies slowly with respect to $x$. 

The rate of turbulent energy production in the mixing layer is directly comparable to the observed luminosity of the mixing layer, assuming that the cooling time of the gas is short, so that the turbulent energy produced is radiated efficiently.\footnote{\citet{KB04} identified the various terms in the internal energy equation relating to the advection of internal plus turbulent energy, the generation of turbulent energy, and its emission via radiation. Since the cooling length in the mixing layer is much less than the advective length, this assumption effectively approximates the energy equation by equating the production of turbulent energy to the emission of radiation. This is similar to the approach taken in classical accretion disc theory \citep[e.g.][]{SS73}.} This is a reasonable assumption in DG~Tau; based on a gas temperature of $10^{4}\textrm{ K}$ and using the cooling function of \citet{SD93}, we determine a cooling time for the IVC of 2.5 yr, which results in a cooling length $\sim 60\textrm{ au}\approx 0.26$~arcsec projected distance for a flow speed of $110\kms$. This cooling length is short compared to the length of the mixing layer, which is $\gtrsim 270$ au (\S\ref{sec:compare-lum}). Furthermore, the fraction of turbulent energy required to dissociate the molecular hydrogen entrained into the shear layer from the ambient wind is small (Appendix \ref{app:H2dissoc}).

We now calculate the rate of turbulent energy production per unit volume at a given point in the mixing layer:
\begin{align}
\frac{\dd\dot{E}_\textrm{turb}(x,\xi)}{\dd V} &= t_{xy}(x,\xi)\frac{\partial \tilde{v}_x}{\partial y} \nonumber\\
 &= t_{xy}(x,\xi)\frac{v_\textrm{jet}}{h(x)} \\
 &= \rho_\textrm{jet} v_\textrm{jet}^3 \frac{h^\prime(x)}{h(x)}\mathcal{F}(\eta,\xi)\textrm{.}
\end{align}
We integrate over $y$ to form an expression for the turbulent energy produced per unit area:
\begin{alignat}{2}
\frac{\dd\dot{E}_\textrm{turb}}{\dd A}(x) &= \int_{y_2}^{y_1} \frac{\dd\dot{E}_\textrm{turb}(x,\xi)}{\dd V}\dd y \\
 &=\rho_\textrm{jet} v_\textrm{jet}^3  h^\prime(x) \mathcal{G}(\eta)\textrm{, where} \label{eq:EturbperA}\\
\mathcal{G}(\eta) &= \frac{2 \eta ^3+3 \eta ^2-6 \eta ^2 \log (\eta )-6 \eta +1}{12 (\eta -1)^4} & \textrm{ for }\eta & \neq 1\textrm{, and} \nonumber\\
 & = \frac{1}{24} & \textrm{ for }\eta & =1\textrm{.} \label{eq:Geta}
\end{alignat}
For the purpose of comparing this model to observations of three-dimensional protostellar jet mixing layers, this is the turbulent energy produced per unit circumference per unit length. Therefore, total turbulent energy production in the layer is calculated multiplying equation (\ref{eq:EturbperA}) by $2\pi R_\textrm{mix}(x)$, where $R_\textrm{mix}(x)$ is the mixing layer radius, and then integrating over the observed mixing layer length, $L$:
\begin{align}
\dot{E}_\textrm{tot} &= \int_0^L 2\pi R_\textrm{mix}(x) \rho_\textrm{jet}v_\textrm{jet}^3h^\prime(x)\mathcal{G}(\eta)\dd x \\
                     &= 2\pi R_\textrm{mix} L \rho_\textrm{jet}v_\textrm{jet}^3h^\prime(x)\mathcal{G}(\eta)\textrm{,} \label{eq:Eturbtotbol}
\end{align}
assuming that $R_\textrm{mix}$ and $h^\prime(x)$ are independent of $x$.

\section{Comparison to Observations}\label{sec:compare}

\subsection{[Fe\ {\sc ii}] 1.644 $\umu$m Shock Modelling}\label{sec:compare-shock}

Our model provides estimates for the bolometric luminosity of a protostellar outflow mixing layer (\S\ref{sec:compare-lum}). However, observations of these outflows are typically made using specific optical and NIR emission lines. Therefore, to compare with observations, we estimate the [Fe\ {\sc ii}] line luminosities for a given total luminosity. To this end, a grid of shock models capable of heating their
post-shock gas to between $2\times 10^4$~K and $6\times 10^4$~K were computed using the {\sc mappings IV} code, version 4.0.1 \citep{SD93,Ae08,NDS12,Nie13,Doe13}, covering both solar abundances \citep{Ae09} and iron-depleted abundances \citep{J09,J13}. We use this grid as a representative model of partially-ionized gas being heated and subsequently cooling; we are not concerned with the shock structure itself. Therefore, we chose pre-shock gas parameters, summarised in Appendix \ref{app:shockmodel-preshock}, that yield densities and temperatures in the post-shock region that are comparable to those expected in the mixing layer. These pre-shock parameters are not intended to be strictly representative of protostellar outflows, nor do we imply that the IVC emission is generated in a single flat-planar shock structure, which would not be a good approximation to the shocks occurring in a turbulent mixing layer.

The shocks were driven into a pre-shock medium with a hydrogen number density of $10^4$~cm$^{-3}$, which approximately matches the conditions in the DG~Tau jet (\S\ref{sec:params-dens}). 
Pre-shock temperatures were selected to explore a wide range of pre-shock ionization parameters (Table \ref{tab:CIE}), based on the {\sc mappings IV} collisional ionization excitation model. Selecting a value of plasma $\beta$, that is, the ratio of thermal to magnetic pressure in the pre-shock material, and a post-shock temperature fixes the shock velocity. 
For the post-shock temperatures given above, the resulting shock velocity is between $20$ and $60$~km s$^{-1}$ for $\beta=1$, in agreement with observations (Paper I).
This computed shock velocity was then used to calculate the properties of the post-shock, cooling gas.
The final [Fe\ {\sc ii}] 1.644 $\umu$m line emission from the shock is then expressed as a fraction of the total of all the line emission 
plus the two-photon emission, which can be a large contributor to total emission in these heated, partially-ionized models.  

\begin{figure}
\includegraphics[width=\columnwidth]{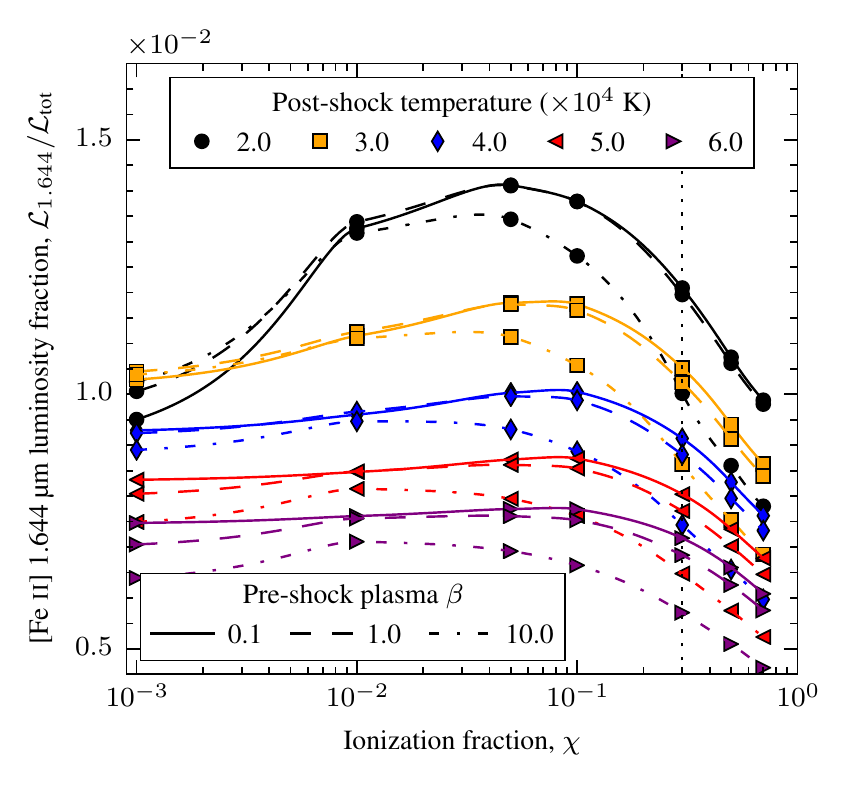}
\caption[Contribution of {[}Fe\ {\sc ii}{]} 1.644 $\umu$m emission to total shock emission]{Contribution of {[}Fe\ {\sc ii}{]} 1.644 $\umu$m line emission to total shock emission from a grid of {\sc mappings IV} models, where the iron abundance is not depleted from solar abundance values and the pre-shock number density $n_\textrm{H}=10^4$. The percentage of shock emission which is radiated as {[}Fe\ {\sc ii}{]} 1.644 $\umu$m line emission is shown as a function of the pre-shock ionization fraction, $\chi$. Models have been computed for a range of post-shock temperatures and pre-shock values of plasma $\beta$ as indicated. Points are joined using a one-dimensional Akima spline interpolation \citep{A70}. The vertical dotted line denotes the known DG Tau jet ionization, $\chi=0.3$.}\label{fig:FeIIlumfrac}
\end{figure}

The results from this model grid are shown in Fig.~\ref{fig:FeIIlumfrac}. Across a range of pre-shock ionizations and plasma $\beta$ (a convenient proxy for magnetic field strength), the fraction of shock luminosity emitted in the [Fe\ {\sc ii}] 1.644 $\umu$m line is $\sim 10^{-2}$. For the purposes of our model, this order-of-magnitude will suffice for comparing with observations.

\begin{figure}
\includegraphics[width=\columnwidth]{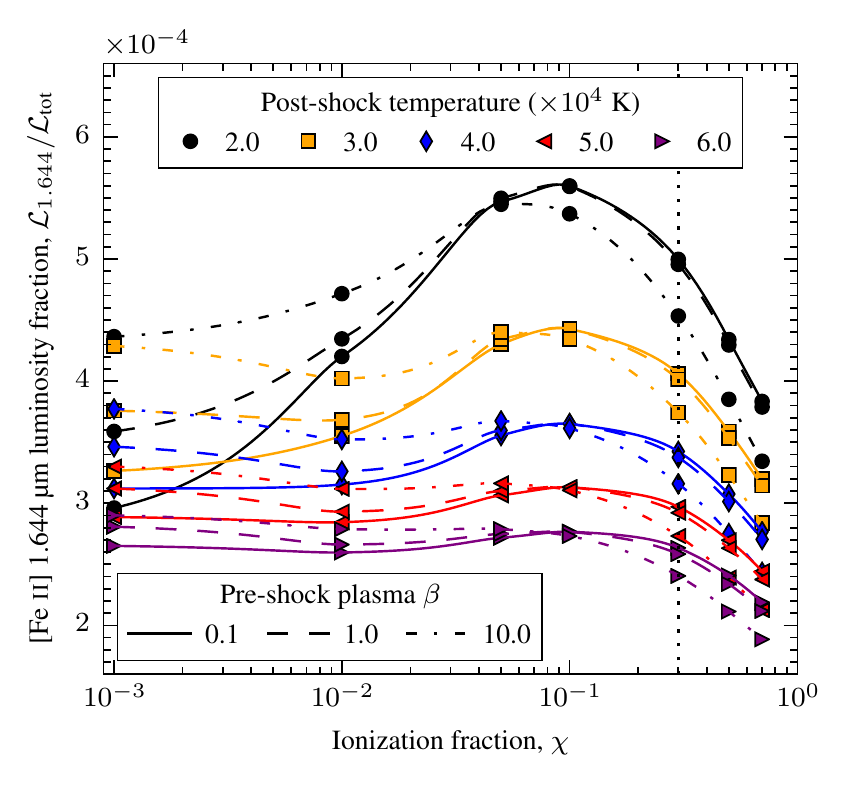}
\caption[Contribution of {[}Fe\ {\sc ii}{]} 1.644 $\umu$m emission to total shock emission for depleted gas]{As for Fig.~\ref{fig:FeIIlumfrac}, but with the pre-shock gas having an iron depletion factor of 100.}\label{fig:FeIIlumfrac-depl}
\end{figure}

The [Fe\ {\sc ii}] total emission affects the model structure, so that the [Fe\ {\sc ii}] 1.644 $\umu$m luminosity does not simply 
scale with the gas phase abundance of iron, but nearly so (see Appendix \ref{app:shockmodel-densdepletion} for details). The cooling fraction in [Fe\ {\sc ii}] 1.644 $\umu$m
with undepleted gas is $\lesssim 1\times 10^{-2}$ (Fig.~\ref{fig:FeIIlumfrac}), and between $2\times 10^{-4}$ and $4\times 10^{-4}$ for the depleted models (Fig.~\ref{fig:FeIIlumfrac-depl}). It is therefore reasonable that the [Fe\ {\sc ii}] 1.644 $\umu$m line emission as a fraction of total 
cooling emission lies between $10^{-2}$ and $10^{-4}$ for a range of iron depletion factors. We parametrize the ratio between bolometric luminosity, $\mathcal{L}_\textrm{tot}$, and [Fe\ {\sc ii}] 1.644 $\umu$m luminosity, $\mathcal{L}_{1.644}$, as\footnote{Note that zero depletion corresponds to a depletion factor of 1.}
\begin{equation}\label{eq:L1644}
\mathcal{L}_{1.644} = \left(\frac{10^{-2}}{\textrm{Fe depletion factor}}\right)\times\mathcal{L}_\textrm{tot}\textrm{.}
\end{equation}

\subsection{Parameters of the DG~Tau Outflow}\label{sec:params}

We summarize the parameters of the DG~Tau jet required as inputs to our model here. The main parameters are the jet density, $\rho_\textrm{jet}$ (\S\ref{sec:params-dens}); the jet velocity, $v_\textrm{jet}$ (\S\ref{sec:params-vel}); the mixing layer growth rate, $h^\prime(x)$ (\S\ref{sec:params-growth}); the mixing layer length, $L$ (\S\ref{sec:params-growth}); and the jet-to-ambient-wind density ratio, $\eta$ (\S\ref{sec:params-densrat}). To utilize our {\sc mappings} {\sc IV} models, we also need to know the iron depletion factor in the DG~Tau IVC (\S\ref{sec:params-depletion}).

\subsubsection{Jet Density}\label{sec:params-dens}

The electron density may be determined from NIR observations through the ratio of the [Fe\ {\sc ii}] emission lines at $1.533\textrm{ }\umu\textrm{m}$ and $1.644\textrm{ }\umu\textrm{m}$. \citet{Pes03} computed a relationship between this line ratio and electron density for a 16-level model of an Fe$^+$ atom. The similar BE99 technique makes use of the ratio of the [S II] emission lines at $6731\textrm{ \AA}$ and $6716\textrm{ \AA}$ in the optical regime \citep{BE99,Me14}. These techniques have been applied to the DG~Tau jet \citep[Paper 1;][]{Be00,A-Ae11,Me14}, yielding a typical electron density $n_\textrm{e}\sim 10^{4}\textrm{ cm}^{-3}$. \citet{Me14} reported higher electron densities, up to $10^6\textrm{ cm}^{-3}$, close to the central star. However, we do not observe significant IVC emission at this position (Paper I, fig.~2.6 therein), so that we use the lower jet density corresponding to the region where we observe a mixing layer.

Our model requires the jet \emph{mass} density as an input. We convert the measured electron density into physical density as in Paper I, \S 4.1.4 therein. The hydrogen density is calculated from $n_\textrm{H} = n_\textrm{e}/\chi_\textrm{e}$, where $\chi_\textrm{e}$ is the ionization fraction of the gas. Although the ionization fraction of the jet appears to vary with position \citep{Me14}, an average ionization fraction of $\chi_e=0.3\pm 0.1$ is a reasonable approximation \citep{Be00}. This yields a hydrogen number density $n_\textrm{H}=3.3\times 10^{4}\textrm{ cm}^{-3}$. The mass density $\rho=1.4mn_\textrm{H}$ for a gas consisting of 90 per cent hydrogen and 10 per cent helium, where $m$ is the atomic mass unit. This calculation leads to a jet mass density $\sim 10^{-19}\textrm{ g cm}^{-3}$, which we use as a fiducial value for our model.

\subsubsection{Jet Velocity}\label{sec:params-vel}

The velocity of the DG~Tau jet varies with time \citep[Paper I;][]{Be02,Pe03b,A-Ae11}, and this is the likely cause of the observed moving shock-excited knots \citep[e.g.][]{Rae90}. The jet velocity is typically measured from the high-velocity peak of line emission \citep{Pe03b,A-Ae11}. More recently, we used multicomponent Gaussian fitting, coupled with a statistical $F$-test, to rigorously separate the two [Fe\ {\sc ii}] 1.644 $\umu$m line-emission components in the approaching DG~Tau outflow (Paper I). These fits show that the high-velocity component of the outflow has a range of velocities from $215\textrm{--}315\kms$ in the 2005 observing epoch. Therefore, we adopt an average jet velocity of $265\kms$ for use in our model.

\subsubsection{Mixing Layer Length and Growth Rate}\label{sec:params-growth}

\begin{figure}
\includegraphics[width=\columnwidth]{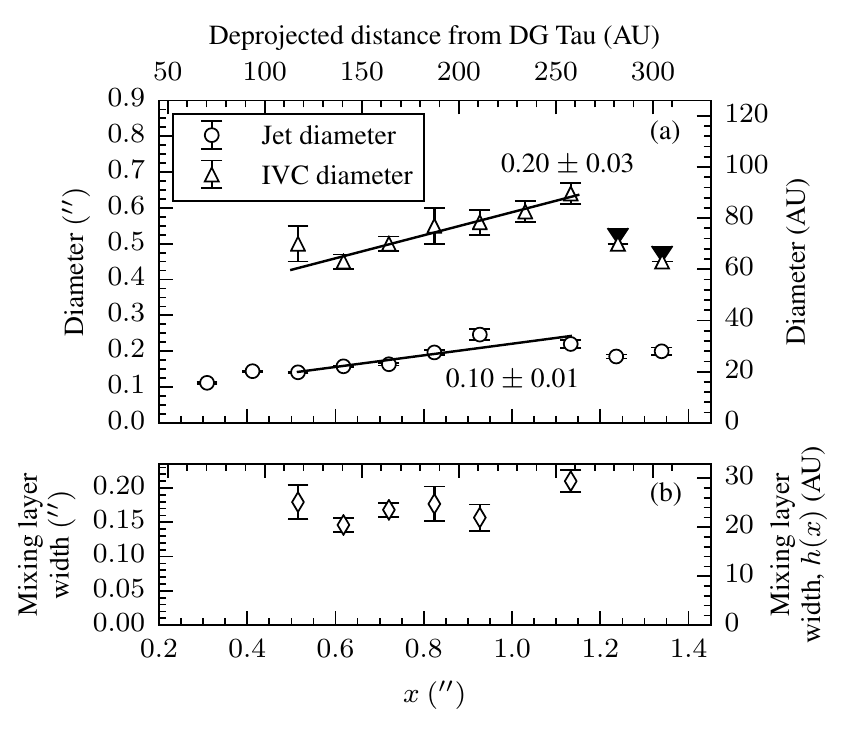}
\caption[Growth rates of the approaching DG~Tau outflow components]{Growth rates of the diameter of the approaching DG~Tau outflow components. (a) Diameter of the approaching jet (circles) and IVC (triangles) as a function of distance from the central star. Jet diameters were measured using cross-jet Gaussian fits to the high-velocity component intensity of [Fe\ {\sc ii}] 1.644 $\umu$m, and are approximately deconvolved from the PSF (Paper I). IVC diameters were measured from direct inspection of images and cross-outflow intensity profiles of the [Fe\ {\sc ii}] 1.644 $\umu$m intermediate-velocity component. Linear fits to the growth of both components over the region $0.5\textrm{--}1.1$~arcsec from the central star are shown as solid lines; the labels indicate the slope of the line (i.e.~the growth rate of the component diameter). (b) Inferred mixing layer widths over the region $0.5\textrm{--}1.1$~arcsec from the central star, as per equation (\ref{eq:hmeasured}).}\label{fig:LVCgrowth}
\end{figure}

The deprojected length of the observed mixing layer, $1.2$~arcsec $\approx 270\textrm{ au}$, can be directly measured from our data (Paper I). The separation of the line emission from the two approaching outflow components allows for the calculation of the mixing layer growth rate. The diameter of the HVC (jet), $D_\textrm{jet}$, was determined by fitting a Gaussian to the [Fe\ {\sc ii}] 1.644 $\umu$m line emission in the cross-jet direction, and approximately deconvolving the width of this Gaussian from the PSF via the formula $D_\textrm{jet}^2=\textrm{FWHM}_\textrm{obs}^2-\textrm{FWHM}_\textrm{PSF}^2$ (Paper I, \S 4.1.4 therein). The diameter of the IVC, $D_\textrm{IVC}$, was measured from both an image of that component (Paper I, fig.~2.6d therein) and cross-outflow [Fe\ {\sc ii}] 1.644 $\umu$m IVC profiles at each downstream position. IVC diameters could not be reliably determined beyond $\sim 1$~arcsec from the central star, due to incomplete line fitting coverage in this region, although conservative lower limits could be inferred.
The component diameters are shown as a function of downstream position in Fig.~\ref{fig:LVCgrowth}(a). The inferred mixing layer width is simply the difference between the observed \emph{radii}, $r_\textrm{jet}$ and $r_\textrm{IVC}$, of the jet and IVC,
\begin{equation}\label{eq:hmeasured}
h(x) = r_\textrm{IVC}-r_\textrm{jet} = \frac{D_\textrm{IVC}-D_\textrm{jet}}{2}\textrm{.}
\end{equation}
The growth rate of the mixing layer is then 
\begin{equation}\label{eq:hprimemeasured}
h^\prime(x)=(D^\prime_\textrm{IVC}-D^\prime_\textrm{jet})/2\textrm{.}
\end{equation}

We determine the growth rate of the mixing layer as follows. We construct linear fits to the lateral growth of both the jet and IVC in the approaching DG~Tau outflow in the 2005 observing epoch over the region $0.5\textrm{--}1.1$~arcsec from the central star (Fig.~\ref{fig:LVCgrowth}a).\footnote{The fits were made using the deprojected distance from the central star and the physical diameter of each outflow component, thereby accounting for the projection of the DG~Tau outflows to the line-of-sight \citep[$38^\circ$;][]{EM98}.} These fits give growth rates of $D^\prime_\textrm{IVC}=0.20\pm 0.03$, and $D_\textrm{jet}^\prime =0.10\pm 0.01$. From equation (\ref{eq:hprimemeasured}), the measured growth rates imply a mixing layer growth rate of $0.05\pm 0.02$. 

The inferred mixing layer width as a function of distance from the central star is shown in Fig.~\ref{fig:LVCgrowth}(b). This is a noisier profile than the individual jet diameters; therefore, it is preferable to determine $h^\prime(x)$ from equation (\ref{eq:hprimemeasured}).

\subsubsection{Jet-to-Ambient Wind Density Ratio}\label{sec:params-densrat}

The density of the jet is well-defined (see \S\ref{sec:params-dens}). In order to estimate the jet-to-ambient-wind ratio, we make approximations to the density of the wider-angle molecular wind in DG~Tau, based on the results of \citet{Te04}. They reported a flow that extended $40\textrm{ au}$ along the outflow axis following deprojection, and $80\textrm{ au}$ across the outflow direction, resulting in a total wind opening angle of $90^\circ$. By considering the $K$-band extinction towards DG~Tau, and the ratio of H$_2$ 1-0 S(1) 2.1218 $\umu$m emitting mass to total H$_2$ mass, they determined a minimum total wind mass in this region of $2.1\times 10^{-8}\textrm{ M}_\odot$. This corresponds to a minimum average wind H$_2$ number density of $4\times 10^{4}\textrm{ cm}^{-3}$, assuming a filling factor of 1 and a conical geometry.

We now make approximations about the flow geometry of this wind in order to determine its density in the entrainment region of the outflow. 
Consider a distance $0.8$~arcsec $\approx 180\textrm{ au}$ from the central star, which is halfway along the observed mixing layer. 
If the wind undergoes no further collimation beyond what is observed in H$_2$ emission, and maintains a conical geometry, it will have a total width of $360\textrm{ au}$ at this position. 
Assuming a constant wind mass-loss rate, $\dot{M}$, and wind velocity, $v$, the wind density, $\rho$, is inversely proportional to the wind radius, $R$, squared:
\begin{equation}
\dot{M}=\rho \pi R^2 v=\textrm{const.}\Rightarrow \frac{\rho_2}{\rho_1}=\frac{R_1^2}{R_2^2}\textrm{.}
\end{equation}
Therefore, an increase in wind radius of a factor $4.5$ would mean a decrease in wind density of a factor of $\sim 20$, resulting in an H$_2$ number density by $\sim 2\times 10^{3}\textrm{ cm}^{-3}$ at $360$ au from the central star. Assuming a gas composition of 90 per cent hydrogen and 10 per cent helium by number density, this results in a mass density of $8.2\times 10^{-21}\textrm{ g cm}^{-3}$, and a jet-to-ambient-wind density ratio of 12.2.

\citet{Te04} notes that their estimates of H$_2$ mass and density in the wider-angle wind are lower limits, given that cold gas may be present in the outflow, and the filling factor of the wind may be less than unity. Hence, our estimate of the jet-to-ambient wind density ratio represents an upper limit to possible values for this parameter. Therefore, whilst we consider it likely that $\eta$ lies between 1 and 10, we have investigated a parameter range of \mbox{$0.1\leq\eta\leq 10$} for completeness.

\subsubsection{Iron Depletion}\label{sec:params-depletion}

The iron depletion in the approaching outflow components from DG~Tau was measured by \citet{A-Ae11} via comparison of [Fe\ {\sc ii}] 1.644 $\umu$m flux to the \mbox{[O I] 6300 \AA} fluxes reported by \citet{LFCD00}. Through comparison with shock wave models \citep{HRP04}, they determined that the iron depletion factor in the approaching DG~Tau outflow is $\sim 3\textrm{--}4$ in gas faster than $-100\kms$ (the jet), and $\sim 10\textrm{--} 12$ for gas at speeds below $-100\kms$ (the IVC).

\citet{A-Ae11} noted that their measurements are tentative, given that they were required to compare [Fe\ {\sc ii}] and [O I] line fluxes obtained $\sim 8\textrm{ yr}$ apart. Furthermore, it is unlikely that the DG~Tau jet would exhibit any iron depletion, as we would expect dust grains to be destroyed by passage through the strong recollimation shock at the base of the approaching outflow (Paper I). However, it is reasonable that the slower, wider-angle outflow components would exhibit iron depletion, as they are launched from wider disc radii and may be less shock-processed \citep{A-Ae11}. Indeed, higher depletion at lower flow velocities has been observed in other YSOs \citep[e.g.~calcium in the HH 111 outflow;][]{Pe09}. The jet iron depletion measurement of \citeauthor{A-Ae11} may be contaminated by IVC emission, given those authors made a simple velocity cut to separate outflow components, rather than using line component fitting \citep[e.g.~Paper I;][]{Le97}. Therefore, we take a range of iron depletion factors, $3\textrm{--}10$, for the DG~Tau approaching IVC.

\subsection{Model Estimates and Comparison for DG~Tau}

We now compare the estimates from our model to our previous observations of the approaching DG~Tau outflow intermediate-velocity component (Paper I). These estimates are based on the outflow parameters for DG~Tau detailed above (\S\ref{sec:params}).

\subsubsection{Mixing Layer Luminosity}\label{sec:compare-lum}

\begin{figure}
\includegraphics[width=\columnwidth]{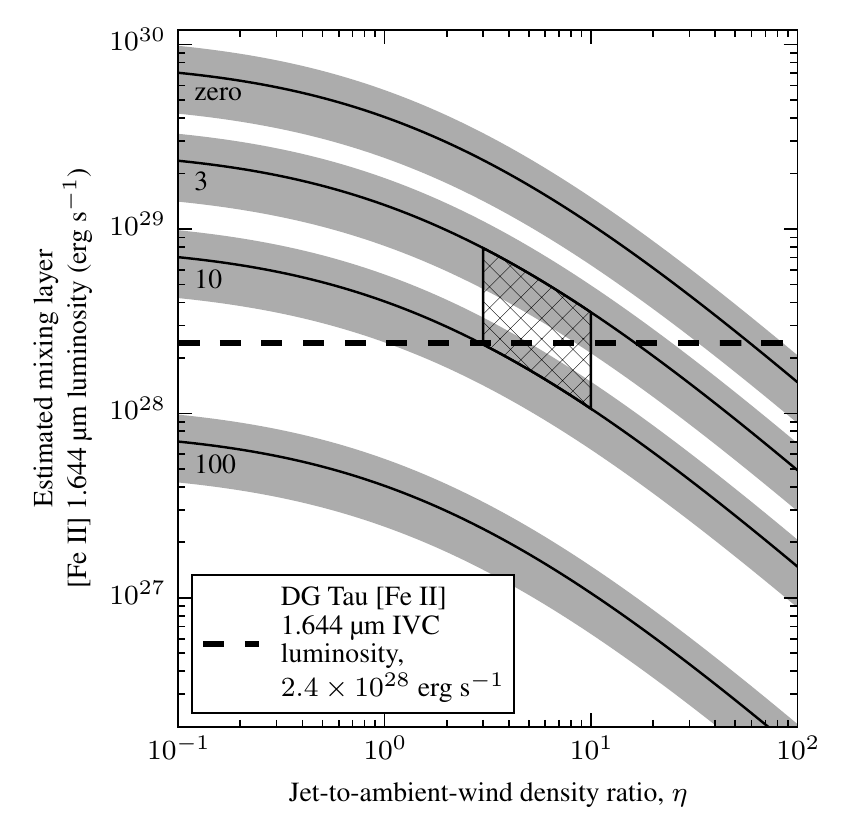}
\caption[Estimates of the DG~Tau mixing layer {[}Fe\ {\sc ii}{]} 1.644 $\umu$m luminosity]{Estimates of the DG~Tau mixing layer {[}Fe\ {\sc ii}{]} 1.644 $\umu$m luminosity, from equations (\ref{eq:L1644}) and (\ref{eq:mixlayerL}). Luminosities are calculated for a range of iron depletion factors and jet-to-ambient-wind density ratios, assuming a jet velocity of $265\kms$ and a jet density of $1.0\times 10^{-19}\textrm{ g cm}^{-3}$. Solid curves show the estimated luminosity for $h^\prime(x)=0.05$; the surrounding greyed regions indicate the range of luminosities at a given iron depletion factor for $0.03\leq h^\prime(x)\leq 0.07$ (\S\ref{sec:params-growth}). Curves are labelled with the corresponding iron depletion factor. The hatched region shows the parameter range applicable to the DG~Tau IVC (iron depletion factor $\sim 3$--$10$; $3\lesssim \eta\lesssim 10$). The thick dashed line shows the observed [Fe\ {\sc ii}] 1.644 $\umu$m luminosity of the DG~Tau approaching IVC, $2.4\times 10^{28}\textrm{ erg s}^{-1}$.}\label{fig:Lpredict}
\end{figure}

The estimated mixing layer luminosity for DG~Tau from our model is shown in Fig.~\ref{fig:Lpredict}. We compute the total mixing layer bolometric luminosity as per \S\ref{sec:model-turb-turbE}, in particular equation (\ref{eq:Eturbtotbol}):
\begin{equation}\label{eq:mixlayerL}
\mathcal{L}_\textrm{mix,tot} = 2\pi R_\textrm{mix} L \rho_\textrm{jet}v_\textrm{jet}^3h^\prime(x)\mathcal{G}(\eta)\textrm{.}
\end{equation}
We estimate the mixing layer radius, $R_\textrm{mix}$, to be $\sim 25\textrm{ au}$ (Fig.~\ref{fig:LVCgrowth}). The mixing layer luminosity is estimated using a jet velocity of $265\kms$ (\S\ref{sec:params-vel}) and a jet density of $10^{-19}\textrm{ g cm}^{-3}$ (\S\ref{sec:params-dens}). We take a range of possible mixing layer growth rates, $h^\prime(x)=0.05\pm 0.02$ (\S\ref{sec:params-growth}). We consider a range of possible iron depletion factors and jet-to-ambient wind density ratios, as these are the least-constrained parameters (\S\S\ref{sec:params-densrat}, \ref{sec:params-depletion}).

We calculate the observed IVC [Fe\ {\sc ii}] 1.644 $\umu$m luminosity from the approaching DG~Tau outflow as follows. We consider every spaxel covering the approaching outflow that was successfully fit with two [Fe\ {\sc ii}] 1.644 $\umu$m emission-line components (Paper I, fig.~2.6 therein). We then calculate the flux from the fitted IVC component in each spaxel, and sum across the entire outflow to produce a total IVC [Fe\ {\sc ii}] 1.644 $\umu$m luminosity of $2.4\times 10^{28}\textrm{ erg s}^{-1}$, assuming a distance to DG~Tau of $140\textrm{ pc}$ \citep{E78}.

Our model estimates for the luminosity of a mixing layer in the approaching DG~Tau outflow is in good agreement with our observations of the approaching IVC. For $3\lesssim\eta\lesssim 10$ (\S\ref{sec:params-dens}), and an iron depletion factor of $\sim 3\textrm{--} 10$, our model estimates a mixing layer luminosity of $(1.1\textrm{--}7.9)\times 10^{28}\textrm{ erg s}^{-1}$. This is a good level of agreement between model and observations, and constitutes a strong indicator that the luminosity of this region of the outflows is driven by turbulent dissipation.

\subsubsection{Rate-of-Change of Mixing Layer Mass Flux}\label{sec:compare-massflux}

In our model, mass enters the mixing layer from both the jet via interception, and from the ambient wind via entrainment. From equation (\ref{eq:Mdot}), the rate at which material enters the mixing layer is
\begin{equation}\label{eq:Mdotprimeexpr}
\frac{\partial \dot{M}}{\partial x}\equiv\dot{M}^\prime = \rho_\textrm{jet}v_\textrm{jet}h^\prime(x)\left( \frac{-\eta+\eta\log(\eta)+1}{(\eta-1)^2} \right)\textrm{.}
\end{equation}
Multiplying by $2\pi R_\textrm{mix}$, where $R_\textrm{mix}\approx 25\textrm{ au}$ as per \S\ref{sec:compare-lum}, gives the entrainment rate per unit length in the outflow direction. Finally, multiplying by the mixing layer length, $L=270\textrm{ au}$ (\S\ref{sec:params-growth}), gives the total mass being gained by the mixing layer at all observed positions (cf.~the calculation of the total turbulent energy production in the mixing layer in \S\S\ref{sec:model-turb-turbE}, \ref{sec:compare-lum}).

\begin{figure}
\includegraphics[width=\columnwidth]{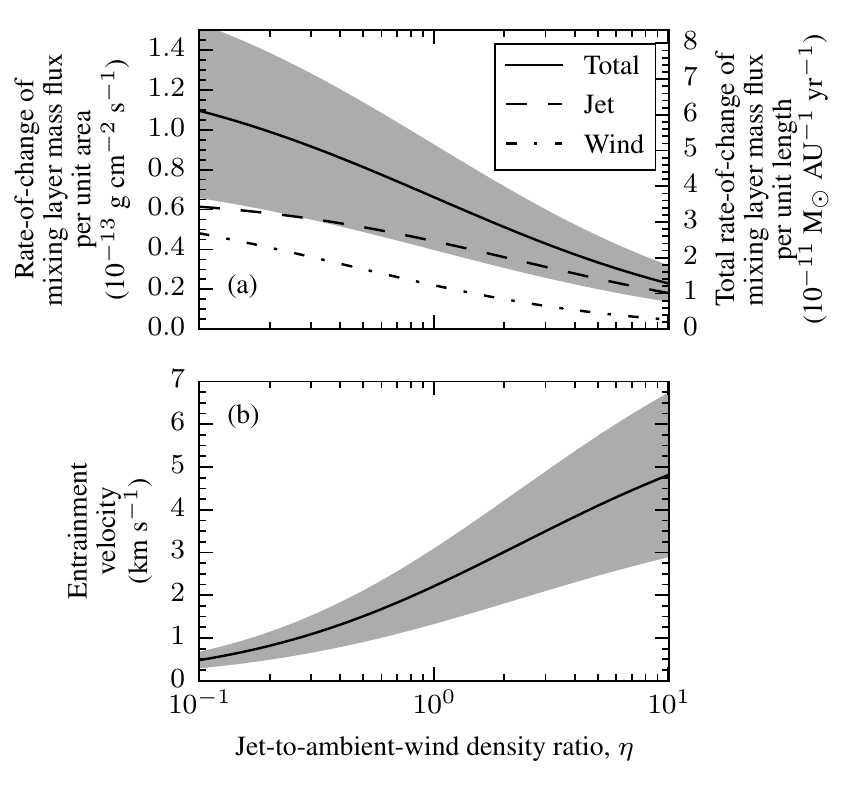}
\caption[Theoretical estimates for DG~Tau mixing layer parameters]{Theoretical estimates for the DG~Tau mixing layer. (a) Estimated rate-of-change of the mixing layer mass flux (solid line), from equation (\ref{eq:Mdotprimeexpr}); contributions from the jet (dashed curve) and ambient wind (dot-dashed curve) are also shown. Rate-of-change of mixing layer mass flux per unit length (right-hand axis) is calculated assuming a mixing layer radius of \mbox{25 au} (\S\ref{sec:compare-massflux}). (b) Estimated entrainment velocity, from equation (\ref{eq:vent}). Estimates are computed using a jet velocity of $265\kms$, a jet density of $10^{-19}\textrm{ g cm}^{-3}$, and a mixing layer growth rate of $0.05$. Greyed regions in both panels show the estimated parameters for a range of mixing layer growth rates, $0.03\leq h^\prime(x) \leq 0.07$.}\label{fig:DGTaupredict}
\end{figure}

An important consistency check is that the total mass being gained by the mixing layer at all positions cannot exceed the combined mass-loss rates of the jet and the wind. Otherwise, the mixing layer would cease to exist at some distance downstream, as it exhausts the mass supply from both sources. For an overdense jet ($1\leq\eta\leq 10$, \S\ref{sec:params-densrat}), our model estimates a total mass gain of $(3.5\textrm{--}10.0)\times 10^{-9}\textrm{ M}_\odot\textrm{ yr}^{-1}$ for the observed mixing layer (Fig.~\ref{fig:DGTaupredict}a). By comparison, the mass-loss rate of the DG~Tau jet is $\sim 5\times 10^{-9}\textrm{ M}_\odot\textrm{ yr}^{-1}$ \citep[Paper I;][]{A-Ae11} from [Fe\ {\sc ii}] emission-line ratios; the total mass-loss rate of all ionized outflow components (jet plus IVC) is $(1\textrm{--}5)\times 10^{-8}\textrm{ M}_\odot\textrm{ yr}^{-1}$ from the VLA data of \citet{Le13}. The mass-loss rate of the molecular wind is lower, $\gtrsim 2.2\times 10^{-9}\textrm{ M}_\odot\textrm{ yr}^{-1}$ \citep{Te04}. However, in the overdense-jet regime, mass interception from the jet is the main contributor to the mass within the mixing layer (Fig.~\ref{fig:Mentcontrib}). Therefore, we conclude that the total mass gain into the mixing layer estimated by our model is less than the combined mass-loss rates of the DG~Tau jet and molecular wind, as required for consistency.

Recently, \citet{Me14} performed an analysis of the DG~Tau approaching outflow using the BE99 technique for determining physical flow parameters from optical line ratios \citep{BE99}. They found that, over the first $0.7$~arcsec of the approaching outflow, the mass outflow rate of the jet (identified as the high-velocity interval, or HVI, in their paper) decreased by $\sim 0.5\textrm{ dex}$. Over the same region, the mass outflow rate of the medium-velocity interval (MVI, which is comparable to our IVC) increased. This is what would be observed if the IVC/MVI represents a turbulent mixing layer which is primarily gaining material from the central jet/HVI.

We compare the observations of \citet{Me14} to our model estimates for the rate-of-change of mixing layer mass flux per unit length, and find them to be consistent. We performed a linear fit to the MVI mass-loss rates of \citet[][fig.~15 therein]{Me14}, and determine an increase in MVI mass-loss rate per unit length of $2\times 10^{-11}\textrm{ M}_\odot\textrm{ au}^{-1}\textrm{ yr}^{-1}$. For the observed parameters of the DG~Tau outflow (\S\ref{sec:params}), our model estimates a rate-of-change in mixing layer mass flux per unit length along the outflow direction of $(1.2\textrm{--}8.6)\times 10^{-11}\textrm{ M}_\odot\textrm{ au}^{-1}\textrm{ yr}^{-1}$, with the lower rate-of-change corresponding to a more overdense jet (Fig.~\ref{fig:DGTaupredict}a).\footnote{For $1\leq\eta\leq 10$, mass entrainment from the ambient wind contributes 20--33 per cent\ of the rate-of-change of mixing layer mass flux (Fig.~\ref{fig:Mentcontrib}), so the mass entrainment per unit length from the ambient wind is $(0.24\textrm{--}2.8)\times 10^{-11}\textrm{ M}_\odot\textrm{ au}^{-1}\textrm{ yr}^{-1}$.} The estimates from our model strongly suggest that the approaching IVC of the DG outflows is consistent with being the signature of a turbulent mixing layer around the central jet.

\subsubsection{Entrainment Velocity and Implied Entrainment Efficiency}\label{sec:compare-vent}

In the jet entrainment models of \citet{CR91} and \citet{RCC95}, material is injected into the mixing layer from the ambient medium/wind with a prescribed entrainment velocity. This velocity is expressed as a fraction of the sound speed in the ambient wind, $c_\textrm{w}$, as it was argued that the ambient wind would be incapable of supplying material at a velocity greater than the sound speed \citep[][however, see \S\ref{sec:disc-prevmodels}]{CR91}. This fraction is defined as the entrainment efficiency, $\epsilon\leq 1$, where the entrainment velocity is written as $v_\textrm{ent}=\epsilon c_\textrm{w}$.

The entrainment velocity of our model is given by equation (\ref{eq:vent}), and is shown as a dimensionless function of $\eta$ in Fig.~\ref{fig:MentG}(a). For jet-to-ambient-wind density ratios $10^{-1}\leq\eta\leq 10^{1}$, the dimensionless entrainment velocity, $v_\textrm{ent}/(v_\textrm{jet}h^\prime(x))$ varies between $\sim 0.005$ and $\sim 0.36$. For a jet velocity of $265\kms$ and  $h^\prime(x) = 0.03$--$0.07$, we estimate a range of entrainment velocities, $0.3\leq v_\textrm{ent}\leq 6.7\kms$. For our inferred values of $\eta\lesssim 10$ and $h^\prime(x)=0.05$, equation (\ref{eq:vent}) estimates an entrainment velocity $\lesssim 5\kms$ (Fig.~\ref{fig:DGTaupredict}b).

Assuming LTE, the H$_2$ 2.1218 $\umu$m emission observed in the approaching DG~Tau outflow has a temperature of $2\times 10^3\textrm{ K}$ \citep{Be08,A-Ae14}. As the ambient wind is not directly observable in the region where entrainment is occurring, we may assume that the wind has cooled somewhat. Therefore, we take an indicative temperature of $10^3\textrm{ K}$, which leads to a sound speed of $2.2\kms$ in the wind for a molecular gas with mean molecular weight $2.3 m_\textrm{H}$.

Direct comparison with the range of entrainment velocities predicted above implies a range of entrainment efficiencies between 0.13 and 3.05 for the full range of possible values for the mixing layer growth rate and jet-to-ambient wind density ratio. Adopting the best-fit value for the mixing layer growth rate, $h^\prime(x)=0.05$, and assuming that the jet is likely to be overdense by up to a factor of 10 (\S\ref{sec:params-densrat}), gives a range of implied entrainment efficiencies from 1.00 to 2.3. We discuss this further in \S\ref{sec:disc-prevmodels}.

\section{Discussion}\label{sec:disc}

\subsection{Comparison with Earlier Models}\label{sec:disc-prevmodels}

\citet{CR91} and \citet{RCC95} utilized laboratory experiments \citep{BE72} to estimate the entrainment efficiency, $\epsilon$, of protostellar jet mixing layers. \citeauthor{RCC95} propose that $\epsilon\sim 0.03$. Furthermore, both \citet{CR91} and \citet{RCC95} claimed that $\epsilon \leq 1$, because the ambient wind should not supply material at greater than the sound speed. However, our model, based only on observable parameters of the protostellar outflows, implies an entrainment efficiency $1\lesssim \epsilon \lesssim 2.5$, in contradiction to earlier models. We argue below that our estimated entrainment efficiency is physically viable.

A detailed analysis of the laboratory experiments of \citet{BE72} is beyond the scope of this paper. However, we make several important points.
First, the experiments of \citeauthor{BE72} concern adiabatic mixing layers. However, both observations \citep[e.g.][]{Be02} and analytical estimates of the mixing layer cooling length (\S\ref{sec:model-turb-turbE}) indicate that the DG~Tau IVC is radiative. There are significant differences between the turbulent mixing layers which form along the jet boundaries in adiabatic and radiative jets, as evidenced by multiple numerical studies. For example, in the adiabatic case, mixing between the jet and the ambient medium results in a large transfer of energy into driving transverse motion of the interacting gas, causing rapid expansion of the jet \citep[e.g.][]{Mie00b}. However, radiative effects assist the jet in forming a turbulent mixing layer consistent with observations by (i) increasing the level of mixing by `breaking' KH-induced waves on the jet surface \citep{DR98}, and (ii) limiting the amount of jet spreading which occurs during mixing, keeping the jet collimated to the large spatial distances observed in, e.g.~HH outflows \citep{Mie98,Mie00b}. This effect is particularly pronounced in overdense jets, such as the DG~Tau jet (\S\ref{sec:params-densrat}).

Secondly, the experiments were conducted for jets of Mach number 1--5, whilst the DG~Tau jet has a (thermal) Mach number of 18--25. More recent experimental work into compressible mixing layers \citep[e.g.][]{SZD00} shows that the mixing layer growth rates appear to approach a constant value for increasing Mach number difference between the flows, suggesting that it may be possible to extract an entrainment efficiency for highly supersonic jets from these experiments. Furthermore, magnetic fields within the jet may lower the effective jet Mach number (Paper I) into the regime investigated by \citet{BE72} and \citet{SZD00}. However, these experiments do not account for magnetic effects and their potential role in turbulence and entrainment, and as discussed above, do not cover the dimensionless parameter space of protostellar jets. Hence, our semi-empirical model is specifically designed to negate a need to rely on laboratory results. We stress that our approach should be considered a complimentary alternative to, rather than a replacement for, the approach taken by, e.g.~\citet{CR91} and \citet{RCC95}.

We now address the issue of entrainment velocities greater than the ambient sound speed.
\citet{CR91} argued that the entrainment velocity of the ambient material must
be subsonic, otherwise extra shocks would form along the mixing layer boundary \citep[e.g.][]{PR88}, conflicting with observations.
However, a supersonic entrainment velocity does not invalidate our model. 
Our work differs from that of \citet{CR91} by not imparting the entrainment velocity on the ambient wind. Rather, the ambient wind is stationary, and material is entrained via the expansion of the shear layer. As ambient material in engulfed by the mixing layer, it is then instantaneously accelerated to $v_\textrm{ent}$. This velocity is supersonic with respect to the ambient wind sound speed, and probably transonic with respect to the shear layer sound speed, so shocks will result; however, these shocks will be standard turbulence internal to the mixing layer, and are therefore consistent with our general approach. There will be no global blunt body-type shock along the shear layer boundary.

In reality, the transition from non-turbulent flow outside the mixing layer to fully turbulent flow within would not be as abrupt as we have modelled here. In particular, the sudden increase in $y$-velocity at the shear layer-ambient wind boundary is likely an artefact from our adoption of a linear shear layer temperature profile; the transition from low-density entraining molecular gas to high-density jet gas is likely to be more gradual. However, even if there are comparable turbulent velocities within the molecular gas being drawn into the layer as in the gas well within the mixing layer, the rate of dissipation per unit volume, $\sim \rho {v^\prime}^3/l_t$, where $\rho$ is the density, $v^\prime$ is the turbulent velocity, and $l_t$ is the turbulent eddy scale size, is lower in the molecular gas than in the mixing layer, because of the lower density of the former.

\subsection{The Extent of the Laminar Jet}\label{sec:disc-laminar}

Jets that undergo lateral entrainment will eventually become completely turbulent, as the inner boundary of the mixing layer expands into the jet and reaches the symmetry axis \citep[e.g.][]{Bi84,DWS85}. This does not appear to occur in the DG~Tau jet within $1.5$~arcsec $\sim 340\textrm{ au}$ of the central star, as is evidenced by the low-velocity-dispersion core of the approaching high-velocity [Fe\ {\sc ii}] 1.644 $\umu$m component (Paper I, fig.~2.6c therein). It is therefore relevant to determine if our model predicts the DG~Tau jet should remain laminar within the NIFS field. Whilst a fully three-dimensional, axisymmetric model is formally required to make this calculation \citep{CR91}, our model provides a useful preliminary exploration.

The jet will become totally turbulent once the jet-mixing layer boundary, $y_1$, reaches the symmetry axis of the jet. The jet-mixing layer boundary position is given by
\begin{equation}
y_1(x) = h(x)\xi_1(\eta) = h^\prime(x)\xi_1(\eta)x\textrm{,}
\end{equation}
assuming $h(x)$ is linear in $x$. The downstream distance at which the jet becomes completely turbulent, $x_\textrm{turb}$, is then simply
\begin{equation}
x_\textrm{turb} = \frac{r_\textrm{jet}}{h^\prime(x)\xi_1(\eta)}\rightarrow \frac{x_\textrm{turb}}{r_\textrm{jet}}=\left(h^\prime(x)\xi_1(\eta)\right)^{-1}\textrm{.}
\end{equation}
For DG~Tau, $h^\prime(x)=0.05$ (\S\ref{sec:params-growth}), $\xi_1(\eta=10)\sim 0.2$ (\S\S\ref{sec:model-turb-profiles}, \ref{sec:params-densrat}), and the maximum observed jet radius $r_\textrm{jet,max}\sim 20\textrm{ au}$ (Fig.~\ref{fig:LVCgrowth}a). The distance from the central star where the DG~Tau jet becomes completely turbulent is then $\sim 2000\textrm{ au}\approx 8.8$~arcsec along the outflow axis, accounting for projection effects. This is well beyond the extent of the NIFS field, in agreement with our earlier observations (Paper I).

\section{Conclusions}\label{sec:concl2}

We have constructed a model of the turbulent lateral entrainment of ambient material by a supersonic, collimated jet (\S\ref{sec:model-turb}). This model aims to explain  the medium-/intermediate-velocity forbidden-line emission that is often seen surrounding YSO jets. The model statistically averages the conservation equations of MHD, and uses directly observable outflow parameters as inputs. Our model calculates the total production, and subsequent dissipation, of energy in a turbulent mixing layer between the jet and the surrounding molecular wind, via calculation of the total turbulent stress within the layer. This allows theoretical estimates of, e.g.~the luminosity and entrainment rate of the mixing layer to be formed.

We computed estimates for the bulk properties of the [Fe\ {\sc ii}] 1.644 $\umu$m IVC observed in the approaching outflow from the YSO DG~Tauri (\S\ref{sec:compare}). We calculated a grid of shock models using the {\sc mappings} {\sc IV} code, to facilitate comparison between the observed [Fe\ {\sc ii}] luminosity of the component, and the estimated bolometric luminosity from our model. Our model accurately estimates the luminosity and rate-of-change of mass flux of the DG~Tau IVC, leading us to conclude that the IVC does indeed represent a turbulent mixing layer between the DG~Tau high-velocity jet, and wider-angle disc wind.

We compared our work with previous models of turbulent entrainment by jets, specifically those of \citet{CR91} and \citet{RCC95}. Unlike the previous models, our adoption of an alternative semi-empirical approach means our model is not dependent upon an `entrainment efficiency' parameter, which must be estimated from laboratory experiments. We argued that the requirement for subsonic `entrainment velocities' from the ambient wind is not necessary in the context of our model. We also estimated the extent of laminar jet flow in DG~Tau (\S\ref{sec:disc-laminar}), although we note that this is simply an illustrative case due to the limitations of our two-dimensional model (see below).

As observed by \citet{CR91}, three-dimensional axisymmetric models of turbulent entrainment by jets are required for definitive analysis of radiative mixing layers. An obvious next step would be to extend the model presented here to three dimensions, and to remove the restriction on the ambient wind having zero streamwise velocity. MHD simulations of turbulent entrainment would also be helpful for characterising the efficiency of the entrainment process, as well as determining the effects of the jet magnetic field on turbulent entrainment.

\section*{Acknowledgements}

Based on observations obtained at the Gemini Observatory, which is operated by the 
Association of Universities for Research in Astronomy, Inc., under a cooperative agreement 
with the NSF on behalf of the Gemini partnership: the National Science Foundation 
(United States), the National Research Council (Canada), CONICYT (Chile), the Australian 
Research Council (Australia), Minist\'{e}rio da Ci\^{e}ncia, Tecnologia e Inova\c{c}\~{a}o 
(Brazil) and Ministerio de Ciencia, Tecnolog\'{i}a e Innovaci\'{o}n Productiva (Argentina).

We are extremely grateful for the support of the NIFS teams at
the Australian National University, Auspace, and Gemini Observatory for their tireless efforts during the instrument integration, commissioning and system verification: Jan Van Harmleen, Peter Young, Mark Jarnyk (deceased), Nick Porecki, Richard Gronke, Robert Boss, Brian Walls, Gelys Trancho, Inseok Song, Chris Carter, Peter Groskowski, Tatiana Paz, John White, and James Patao. Thanks must also go to Tracy Beck for her support during the commissioning and observations.

MW acknowledges the generous travel support from Academia Sinica to attend the conference Star Formation Through Spectroimaging at High Angular Resolution in July 2011, which provided useful information for this study. This work was supported by the Australian Research Council through Discovery Project Grant DP120101792 (R.~Salmeron).

\bibliographystyle{mnras}
\bibliography{library}

\onecolumn
\appendix

\section{Supplementary Calculations}

\subsection{Characteristic Equations of MHD}\label{app:suppl-chareqn}

For magnetohydrodynamic fluids with density $\rho$, velocity $v$, pressure $p$, viscous stress tensor $t_{ij}^v$, magnetic field $\vec{B}$ immersed in a gravitational potential $\phi_\textrm{G}$, the equation of mass continuity can be written in Cartesian coordinate notation thus:
\begin{equation}\label{eq:eoc-gen}
\frac{\partial\rho}{\partial t}+\frac{\partial(\rho v_i)}{\partial x_i} = 0\textrm{.}
\end{equation}
Similarly, the equation of momentum conservation is written as
\begin{equation}\label{eq:com-gen}
\frac{\partial(\rho v_i)}{\partial t}+\frac{\partial (\rho v_i v_j)}{\partial x_j} = -\rho\frac{\partial\phi_\textrm{G}}{\partial x_i}-\frac{\partial p}{\partial x_i}+\frac{\partial t_{ij}^B}{\partial x_j}+\frac{\partial t_{ij}^v}{\partial x_j}\textrm{} 
\end{equation}
\citep{KB04}.

Time-averaging equation (\ref{eq:eoc-gen}) is trivial, yielding equation (\ref{eq:eocmixing}). Averaging the momentum conservation equation, equation (\ref{eq:com-gen}), is more complex. The viscous stress tensor, $t_{ij}^v$, is disregarded, as it is unimportant to the transfer of momentum on large scales. The Reynolds stress tensor, $t_{ij}^R$, appears in equation (\ref{eq:commixing}) as a result of averaging the second term on the left-hand side of equation (\ref{eq:com-gen}). This may then be combined with the magnetic stress tensor, $t_{ij}^B$, to form a single term encapsulating the total stress in the system, $t_{ij}=t_{ij}^R+t_{ij}^B$.

\subsection{Order of Magnitude Calculations}\label{app:oom}

In this section, we conduct an order of magnitude analysis of the characteristic equations in \S\ref{sec:chareqn}. Our aims in this section are:
\begin{enumerate}
\item Justify the dropping of the $t_{xx}$ term from equation (\ref{eq:eommixing});
\item Demonstrate the constancy of total (thermal plus turbulent) pressure across the mixing layer.
\end{enumerate} 

Consider the orders of magnitude of the terms in the equation of mass continuity, equation (\ref{eq:eocmixing}), as shown by the expressions beneath the braces:
\begin{equation}
\underbrace{\frac{\partial (\bar{\rho}\tilde{v}_x)}{\partial x}}_{\bar{\rho}{\tilde{v}_x}/L}+\underbrace{\frac{\partial (\bar{\rho}\tilde{v}_y)}{\partial y}}_{ \bar{\rho}{\tilde{v}_y}/h } = 0 \Rightarrow \tilde{v}_y\sim \frac{h}{L}\tilde{v}_x\textrm{.}
\end{equation}
It can be shown from our definition of $\xi$, equation (\ref{eq:zeta}), and the transformed equation of mass continuity, equation (\ref{eq:eocxxi}), that $h/L \sim h^\prime$.

It then follows that the orders of magnitude in the equation of conservation of streamwise momentum, equation (\ref{eq:eommixing}), are:
\begin{equation}\label{eq:eomxmag}
\underbrace{\frac{\partial(\bar{\rho}\tilde{v}_x^2)}{\partial x}}_{\rho v^2/L}
+ \underbrace{\frac{\partial( \bar{\rho}\tilde{v}_x\tilde{v}_y )}{\partial y}}_{\rho v^2/L}
= -\underbrace{\frac{\partial\langle \rho v_x^{\prime 2} \rangle}{\partial x}}_{\textrm{(R1): }\rho v_\textrm{t}^2/L}
- \underbrace{\frac{\partial \langle \rho v_x^\prime v_y^\prime \rangle}{\partial y}}_{\textrm{(R2): }\rho v_\textrm{t}^2/h}
+ \underbrace{\frac{\partial}{\partial x}\left( \frac{\langle B_x^{\prime 2}\rangle - \langle B_y^{\prime 2}\rangle - \langle B_z^{\prime 2}\rangle}{8\pi} \right)}_{\textrm{(R3): }B_\textrm{t}^2/8\pi L}
+ \underbrace{\frac{\partial}{\partial y}\left( \frac{\langle B_x^\prime B_y^\prime \rangle}{4\pi} \right)}_{\textrm{(R4): }B_\textrm{t}^2/4\pi h}\textrm{.}
\end{equation}
Note that we have expanded the stress terms, and for the purposes of expressing orders of magnitude, have replaced $\bar{\rho}$ with $\rho$, $\tilde{v}_x$ with $v$, and have denoted the generic turbulent velocity, typically $v^\prime$, as $v_\textrm{t}$ for clarity. Because $h^\prime \ll 1 \Rightarrow L \gg h$, terms (R2) and (R4) of equation (\ref{eq:eomxmag}) are the most important terms on the right-hand side. Therefore, the $t_{xx}$ component of the turbulent stress (encapsulated in the third term on the right-hand side of equation \ref{eq:eomxmag}) is unimportant (\S\ref{sec:chareqn}).

We now must determine which of terms (R2) and (R4) in equation (\ref{eq:eomxmag}) is most important. The observed FWHM of the blueshifted [Fe\ {\sc ii}] 1.644 $\mu$m emission-line component from DG~Tau is $\gtrsim 60\kms$ (Paper I), indicating that the turbulent velocity in the mixing layer, $v_\textrm{t}$, is of that order. Recalling that we assume that the magnetic field is dominated by its turbulent component within the mixing layer (\S\ref{sec:chareqn}), we may compute the turbulent Alfv\'{e}n velocity, $v_{\textrm{A,t}}=B/\sqrt{4\pi}\rho$, using the inferred magnetic field strength in the DG~Tau jet, which is in the range 30-100~$\mu$G \citep{LFCD00,Ae14}. Assuming an equipartition magnetic field of $\sim 60$~$\mu$G, this yields a turbulent Alfv\'{e}n velocity of $\sim$1.7$\kms \ll v_\textrm{t}$; hence, the turbulent velocity term, (R2), is dominant. This implies that
\begin{equation}
\frac{\rho v^2}{L}\sim \frac{\rho v_t^2}{h} \Rightarrow v_t^2\sim\frac{h}{L}v^2\textrm{.}
\end{equation}

Consider the equation for the conservation of transverse momentum, i.e., the $i=y$ expansion of equation (\ref{eq:commixing}):
\begin{equation}\label{eq:eomymag}
\underbrace{\frac{\partial ( \bar{\rho}\tilde{v}_x\tilde{v}_y)}{\partial x}}_{\textrm{L1: }(\rho v^2/L)(h/L)}
+ \underbrace{\frac{\partial (\bar{\rho}\tilde{v}_y^2)}{\partial y}}_{\textrm{L2: }(\rho v^2/L)(h/L)}
= \underbrace{-\frac{\partial \bar{p}}{\partial y}}_{\textrm{(R1): }p/h}
- \underbrace{\frac{\partial \langle \rho v^\prime_x v^\prime_y \rangle}{\partial x}}_{\textrm{(R2): }(\rho v^2/L)(h/L)}
- \underbrace{\frac{\partial \langle \rho v^{\prime 2}_y \rangle}{\partial y}}_{\textrm{(R3): }\rho v^2/L}
+ \underbrace{\frac{\partial}{\partial x}\left(\frac{\langle B_x^\prime B_y^\prime\rangle}{4\pi}\right)}_{\textrm{(R4): }B_t^2/4\pi L}
+ \underbrace{\frac{\partial}{\partial y}\left( \frac{\langle B_y^{\prime 2}\rangle - \langle B_x^{\prime 2}\rangle - \langle B_z^{\prime 2}\rangle}{8\pi} \right)}_{\textrm{(R5): }B_t^2/8\pi h}\textrm{.}
\end{equation}
On the right-hand side of this equation, terms (R3) and (R5) dominate terms (R2) and (R4); then, as in equation (\ref{eq:eomxmag}), the turbulent velocity term (R3) dominates the turbulent magnetic field term (R5). It can also be seen that term (R3) also dominates both of the terms on the left-hand side of the equation because $h/L\ll 1$. Therefore, the two dominant terms of equation (\ref{eq:eomymag}) are terms (R1) and (R3).

Discarding the unimportant terms of equation (\ref{eq:eomymag}), and integrating over $y$, yields the `total' pressure,
\begin{equation}\label{eq:eomymagint}
p_\textrm{tot} \equiv \bar{p} + \langle \rho v_y^{\prime 2} \rangle = \textrm{const.}
\end{equation}
This relationship allows us to define a new quantity, $\mathcal{W}$:
\begin{equation}\label{eq:W}
\mathcal{W} \equiv \tilde{T}_\textrm{thermal} + \frac{\mu m \langle \rho v_y^{\prime 2}\rangle}{k \bar{\rho}}\textrm{,}
\end{equation}
which is related to density and total pressure thus:
\begin{equation}
p_\textrm{tot} = \frac{\bar{\rho}k}{\mu m}\mathcal{W}\textrm{.}
\end{equation}
We approximate $\mathcal{W}$ as being linear across the mixing layer, with bounds given by equation (\ref{eq:Tprofile}). This is a reasonable approximation, given:
\begin{enumerate}
\item The constancy of total pressure across the mixing layer (equation \ref{eq:eomymagint});
\item The fact that turbulent velocities (that is, velocity dispersion), are present in at least both the mixing layer and the jet (Paper I, fig.~6 therein).
\end{enumerate}

\subsection{Dimensionless Functions}\label{app:suppl-dimless}

The dimensionless function $\mathcal{D}(\eta,\xi)$ in the mixing layer transverse velocity profile, equation (\ref{eq:vy}), is given by
\begin{align}
\mathcal{D}(\eta, \xi) &= \int_{\xi_1}^\xi\xi^\prime\frac{\df}{\df\xi^\prime}\left(\frac{U(\xi^\prime)}{\eta +(1-\eta)S(\xi^\prime)}\right)\dd\xi^\prime \label{eq:calD}\\
 &= \frac{\eta}{(\eta-1)^2}\left\{ (\eta -1) \left[\frac{\xi }{(\eta -1)(\xi_1-\xi )+1}-\xi _1\right] +\log \left[(\eta -1)(\xi_1-\xi )+1\right]\right\}\textrm{.} \label{eq:calDclosed}
\end{align}
The dimensionless function $\mathcal{F}(\eta,\xi)$ in the mixing layer transverse turbulent stress profile, equation (\ref{eq:txy}), is given by
\begin{align}
\mathcal{F}(\eta,\xi) &= \int_{\xi_1}^\xi -\xi^\prime\frac{\xi^\prime-\xi_1+1}{\eta+(1-\eta)(\xi^\prime-\xi_1+1)}+\mathcal{D}(\eta,\xi^\prime)\dd\xi^\prime\textrm{.}\label{eq:calT} \\
 & = \frac{1}{2(\eta-1)^3} \Big\{2 \eta  \left[(1-\eta)(\xi_1-\xi)-1\right]\log \left[(\eta-1)(\xi_1-\xi)+1\right] \nonumber\\
 & \phantom{=\frac{1}{2(\eta-1)^3}[} - \left[(\eta-1)(2\eta-1)\xi_1-\eta(\xi-2)+\xi \right] (\eta -1) \left(\xi -\xi _1\right)\Big\}\textrm{.}\label{eq:calTclosed}
\end{align}

\subsection{Mixing Layer Transverse Velocity and Turbulent Stress Profiles}\label{app:suppl-profiles}

The transverse velocity profile across the mixing layer, $\tilde{v}_y(\eta,\xi)$, may be found by substituting the expression for $\xi_1(\eta)$, equation (\ref{eq:xi1eta}), into equation (\ref{eq:vy}):
\begin{align}
\tilde{v}_y(\eta,\xi) & = \frac{v_\textrm{j}h^\prime(x)\eta}{4(\eta-1)^6}\Bigg\{ \left[\eta ^2(2\log(\eta)-1)+1\right] \left[3 \eta ^2-2 \eta ^2 \log (\eta )+2 (\eta -1)^3 \xi -4 \eta +1\right] \nonumber\\
                      &\phantom{=\frac{v_\textrm{j}h^\prime(x)\eta}{4(\eta-1)^6}\Bigg\{ } -2 (\eta -1)^2 \left[\eta ^2-2 \eta ^2 \log (\eta )+2 (\eta -1)^3 \xi -1\right] \log \left(-\frac{\eta ^2(1-2\log(\eta))+2 (\eta -1)^3 \xi -1}{2 (\eta -1)^2}\right)\Bigg\}\textrm{.}\label{eq:vyfull}
\end{align}
The turbulent stress profile across the mixing layer, $t_{xy}(\eta,\xi)$, may be found in the same way, using equation (\ref{eq:txy}):
\begin{align}
t_{xy}(\eta,\xi) & = \frac{\rho_\textrm{j}v_\textrm{j}^2h^\prime(x)}{8(\eta-1)^7}\Bigg\{ 4\eta\Big[(\eta -1)^2 \left[\eta ^2 (1-2 \log (\eta ))+2 (\eta -1)^3 \xi -1\right] \nonumber\\
                 &\phantom{=\frac{\rho_\textrm{j}v_\textrm{j}^2h^\prime(x)}{8(\eta-1)^7}\Bigg\{ 4\eta\Bigg[} \times\log \left(-\frac{\eta ^2(1-2\log (\eta ))+2 (\eta -1)^3 \xi -1}{2 (\eta -1)^2}\right) \nonumber\\
			     &\phantom{=\frac{\rho_\textrm{j}v_\textrm{j}^2h^\prime(x)}{8(\eta-1)^7}\Bigg\{ 4\eta\Bigg[} +\eta  \log (\eta ) \Big(\eta  \left(-4 \eta ^2-2 (\eta -1)^3 \xi +7 \eta -4\right) +(2 \eta -1) \eta ^2 \log (\eta )+1\Big)\Big] \nonumber\\
			     &\phantom{=\frac{\rho_\textrm{j}v_\textrm{j}^2h^\prime(x)}{8(\eta-1)^7}\Bigg\{ }+(\eta-1)^2 \Big[ 4 (\eta -1)^4 \xi ^2+4 \eta  (\eta +1) (\eta -1)^2 \xi +\eta  (\eta  (6 \eta -5)+4)-1\Big]\Bigg\}\textrm{.} \label{eq:txyfull}
\end{align}

\subsection{Calculation of the Mass Entrainment Rate}\label{app:suppl-Mentrate}

The mass entrainment rate is given by taking the $x$-derivative of equation (\ref{eq:Mdotent}),
\begin{align}
\dot{M}^\prime_\textrm{ent} & = \frac{\textrm{d}}{\textrm{d}x}\int_{\xi_2}^{\xi_1}\rho(\xi)v_x(x,\xi)h(x)\dd\xi - \rho_\textrm{j}v_\textrm{j}h^\prime(x)\xi_1(\eta) \label{eq:Mdotprimeent}\\
 						    & = \bar{\rho}(\xi_1)\tilde{v}_x(\xi_1)h(x)\frac{\df\xi_1}{\df x}-\bar{\rho}(\xi_2)\tilde{v}_x(\xi_2)h(x)\frac{\df\xi_2}{\df x} \nonumber\\
 						    & \phantom{=}+\int_{\xi_2}^{\xi_1}\frac{\df}{\df x}\left[\rho(\xi)v_x(x,\xi)h(x)\right]\dd\xi-\rho_\textrm{j}v_\textrm{j}h^\prime(x)\xi_1(\eta)\textrm{.}\label{eq:Mdotprimeentexp}
\end{align}
The first two terms of the above are zero in $(x,\xi)$-space. The transformed equation of continuity, equation (\ref{eq:eocxxi}), may be written as
\begin{equation}
\frac{\partial}{\partial x}\left( \bar{\rho}\tilde{v}_xh(x)\right) = h^\prime(x)\frac{\partial}{\partial \xi}\left(\xi\bar{\rho}\tilde{v}_x\right) - \frac{\partial}{\partial \xi}\left(\bar{\rho}\tilde{v}_y\right)\textrm{,}
\end{equation}
which reduces equation (\ref{eq:Mdotprimeentexp}) to
\begin{equation}
\dot{M}^\prime_\textrm{ent}=h^\prime(x)\left[\rho_\textrm{j}v_\textrm{j}\xi_1-\rho_\textrm{w}v_\textrm{w}\xi_2 \right]-\rho_\textrm{J}\tilde{v}_y(x,\xi_1)+\rho_\textrm{w}\tilde{v}_y(x,\xi_2)-\rho_\textrm{j}v_\textrm{j}\xi_1h^\prime(x)\textrm{.}
\end{equation}
Most of these terms are zero, or cancel, leaving
\begin{align}
\dot{M}^\prime_\textrm{ent} & = \rho_\textrm{w}\tilde{v}_y(x,\xi_2) \\
                            & = \rho_\textrm{w}v_\textrm{ent}\textrm{ by definition.}
\end{align}

\section{Dissociation of Entrained Molecular Hydrogen Within the Mixing Layer}\label{app:H2dissoc}

The calculation presented in this paper does not account for the energy required to dissociate hydrogen molecules (H$_2$) that are entrained into the mixing layer. In this Appendix, we show that the energy required to dissociate the entrained H$_2$ is of the order of 1 per cent of all turbulent energy produced in the mixing layer, and hence has a negligible effect on the comparison to the observed mixing layer luminosity in \S\ref{sec:compare-lum}.

The mass of molecular material being entrained into the mixing layer, $\dot{M}_\textrm{ent}$, may be computed in a similar fashion to the total rate-of-change of mixing layer mass flux (\S\ref{sec:compare-massflux}), by multiplying the local instantaneous mass entrainment rate ($\dot{M}^\prime_\textrm{ent}$, equation \ref{eq:Mdotent}) by the mixing layer length $L$, and by $2\pi R_\textrm{mix}$, where $R_\textrm{mix}$ is an approximation of the mixing layer radius. Therefore, the total mass of molecular wind material entering the mixing layer at any one time is
\begin{equation}\label{eq:Menttot}
M_\textrm{ent} = \underbrace{\frac{\rho_\textrm{jet}}{\eta}}_{\rho_\textrm{w}}
\underbrace{v_\textrm{jet}h^\prime(x)\frac{\eta(\eta^2-2\eta\log(\eta)-1)}{2(\eta-1)^3}}_{v_\textrm{ent}}
2 \pi R_\textrm{mix} L \textrm{.}
\end{equation}

\begin{figure}
\centering
\includegraphics[scale=1.0]{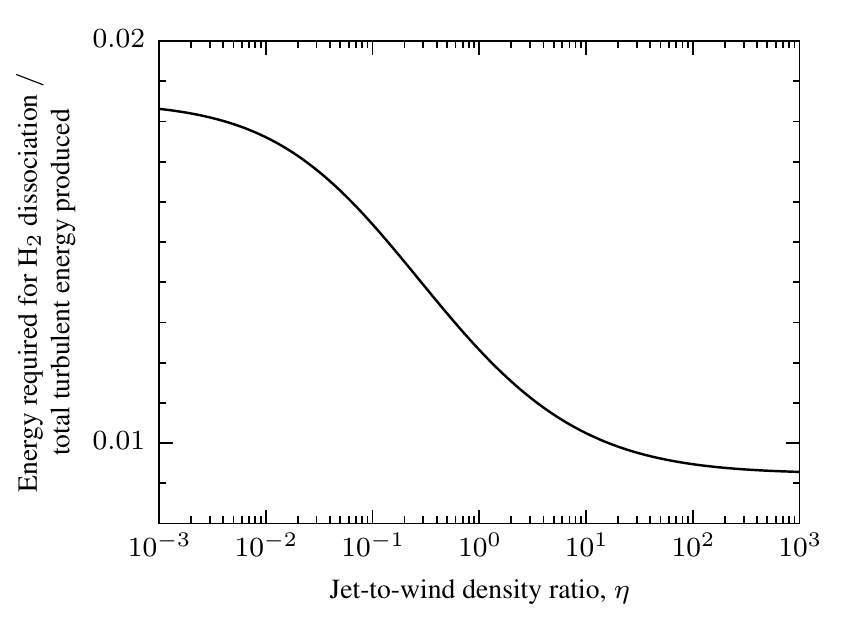}
\caption{The ratio of the energy required to dissociate all the H$_2$ molecules entrained into the turbulent mixing layer to the total turbulent energy produced within the mixing layer. The ratio is plotted as a function of the jet-to-ambient-wind density ratio, $\eta$. The ratio is computed assuming a jet velocity of $265\kms$, a mixing layer growth rate of 0.05, and a jet density of $1.0\times 10^{-19}$~g cm$^{-3}$, as for DG~Tau (\S\ref{sec:params}).}\label{fig:H2dissoc}
\end{figure}

The bond dissociation energy of molecular hydrogen is 4.52 eV per molecule~$=2.2\times 10^{12}$~erg g$^{-1}$ \citep{chem_BE03}. The total energy required to dissociate all of the molecular hydrogen present in the mixing layer may then be calculated by multiplying this value by equation (\ref{eq:Menttot}). The ratio of this required energy is compared to the total turbulent energy produced within the mixing layer, equation (\ref{eq:Eturbtotbol}), in Fig.~\ref{fig:H2dissoc}. The energy required to dissociate the entrained molecular material is of the order of 1 per cent of the total turbulent energy produced. Given that our shock models indicate that at most 1 per cent of the total turbulent energy produced is radiated away as [Fe\ {\sc ii}] 1.644 $\mu$m emission, the effect on our predicted [Fe\ {\sc ii}] mixing layer luminosity is of the order of 0.01 per cent, i.e.~negligible.

\section{Shock Modelling Supplementary Information}\label{app:shockmodel}

\subsection{Pre-Shock Gas Parameters}\label{app:shockmodel-preshock}

A collisional ionization equilibrium (CIE) model was used to determine the relative abundances of various ions in the pre-shock gas. The results of these models are shown in Table \ref{tab:CIE}. Conversion of these fractions to number densities was then achieved by multiplying by the relevant elemental abundances (\citealt{Ae09} for solar abundances, and \citealt{J09,J13} for the depleted case). The weak dependence of [Fe\ {\sc ii}] 1.644 $\umu$m luminosity fraction on pre-shock ionization fraction (Figs.~\ref{fig:FeIIlumfrac}, \ref{fig:FeIIlumfrac-depl}) can be seen in this Table; the relative fraction of iron in the singly ionized state does not vary significantly with hydrogen ionization.

\ctable[
caption = {Pre-shock ionization fractions based on collisional ionization equilibrium models.},
cap = {Pre-shock ionization fractions},
label = {tab:CIE},
]{r@{\qquad}ccc@{\qquad}ccc@{\qquad}ccc}{
\tnote[a]{Shown are the fractions of each species in the neutral (I), singly ionized (II) and doubly ionized (III) states as a function of hydrogen ionization fraction, $\chi$.}
\tnote[b]{Temperatures determined from the collisional ionization equilibrium model.}
}{
\hline
 & \multicolumn{3}{c@{\qquad}}{$\chi = 0.01$\tmark[a]} & \multicolumn{3}{c@{\qquad}}{$\chi = 0.05$} & \multicolumn{3}{c@{\qquad}}{$\chi = 0.1$} \\
 & \multicolumn{3}{c@{\qquad}}{T = $1.167\times 10^4$ K\tmark[b]} & \multicolumn{3}{c@{\qquad}}{T = $1.253\times 10^4$ K} & \multicolumn{3}{c@{\qquad}}{T = $1.312\times 10^4$ K} \\
Species & I & II & III & I & II & III & I & II & III \\
\hline
H & 0.991 & 0.009 & --- &  0.951 & 0.049 & --- & 0.899 & 0.101 & --- \\
He & 1.000 & 0.000 & 0.000 & 1.000 & 0.000 & 0.000 & 0.997 & 0.003 & 0.000 \\
Fe & 0.368 & 0.631 & 0.001 & 0.291 & 0.706 & 0.003 & 0.247 & 0.746 & 0.007 \\
\hline
\hline
 & \multicolumn{3}{c@{\qquad}}{$\chi = 0.3$} & \multicolumn{3}{c@{\qquad}}{$\chi = 0.5$} & \multicolumn{3}{c@{\qquad}}{$\chi = 0.7$} \\
 & \multicolumn{3}{c@{\qquad}}{T = $1.445\times 10^4$ K} & \multicolumn{3}{c@{\qquad}}{T = $1.549\times 10^4$ K}&\multicolumn{3}{c@{\qquad}}{T = $1.671\times 10^4$ K} \\
 & I & II & III & I & II & III & I & II & III \\
 \hline 
H & 0.699 & 0.301 & --- & 0.502 & 0.498 &    ---   &  0.301 & 0.699 & --- \\
He & 1.000 & 0.000 & 0.000 & 1.000 & 0.000 & 0.000 & 1.000 & 0.000 & 0.000 \\
Fe & 0.170 & 0.797 & 0.033 & 0.125 & 0.789 & 0.085 & 0.083 & 0.712 & 0.205\\
\hline
}

\subsection{Effect of Varying Pre-Shock Density and Iron Depletion}\label{app:shockmodel-densdepletion}

\begin{figure}
\centering
\includegraphics[scale=1.0]{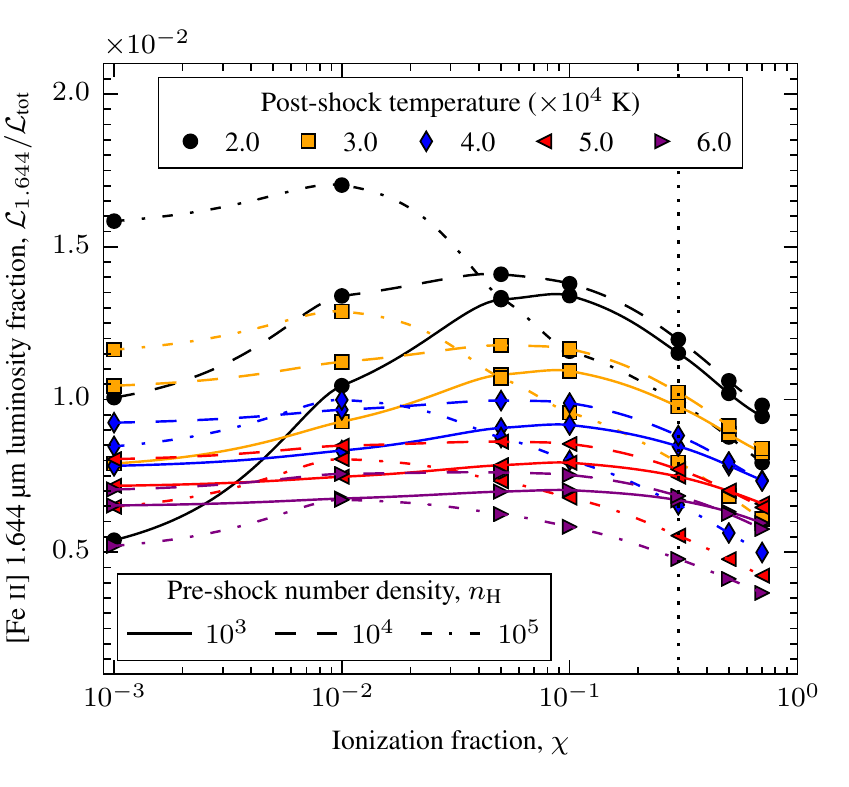}
\caption[Contribution of {[}Fe\ {\sc ii}{]} 1.644 $\umu$m emission to total shock emission]{Contribution of {[}Fe\ {\sc ii}{]} 1.644 $\umu$m line emission to total shock emission from a grid of {\sc mappings IV} models, where the iron abundance is not depleted from solar abundance values and the pre-shock hydrogen ionization $\chi=0.3$. The percentage of shock emission which is radiated as {[}Fe\ {\sc ii}{]} 1.644 $\umu$m line emission is shown as a function of the pre-shock ionization fraction, $\chi$. Models have been computed for a range of post-shock temperatures and pre-shock number densities as indicated. Points are joined using a one-dimensional Akima spline interpolation \citep{A70}.  The vertical dotted line denotes the known jet ionization, $\chi=0.3$.}\label{fig:FeIIlumfrac-dens}
\end{figure}

\begin{figure}
\centering
\includegraphics[scale=1.0]{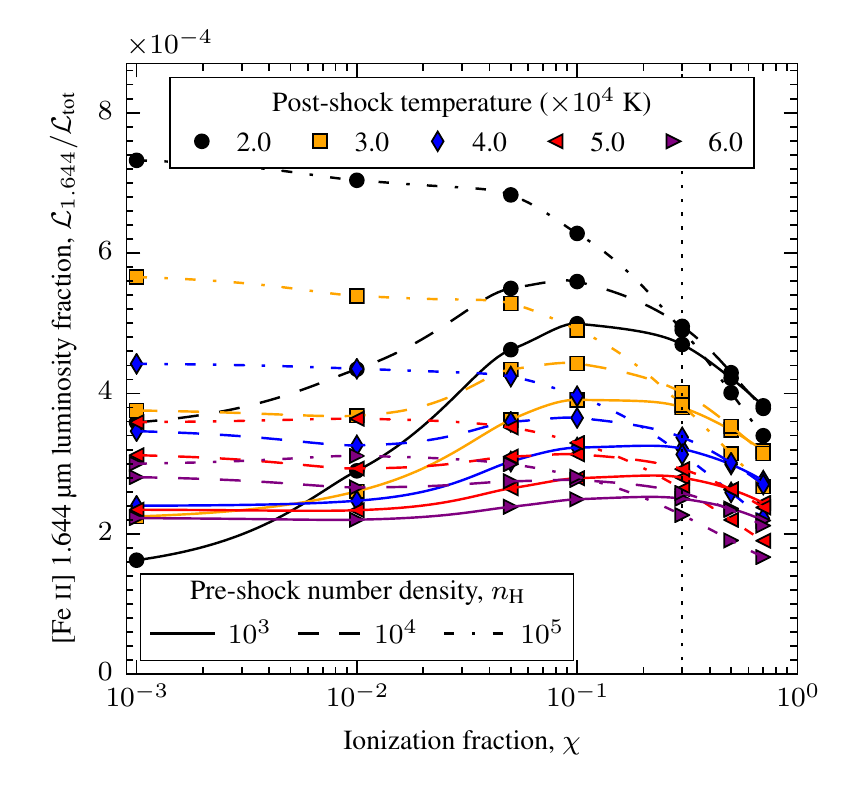}
\caption[Contribution of {[}Fe\ {\sc ii}{]} 1.644 $\umu$m emission to total shock emission]{As for Fig.~\ref{fig:FeIIlumfrac-dens}, but with an iron depletion factor of 100.}\label{fig:FeIIlumfrac-dens-depl}
\end{figure}

Shown in Figs.~\ref{fig:FeIIlumfrac-dens} and \ref{fig:FeIIlumfrac-dens-depl} are the dependencies of the fraction of total emission generated by [Fe\ {\sc ii}] 1.644 $\umu$m line emission as a function of pre-shock ionization and pre-shock number density (cf.~Figs.~\ref{fig:FeIIlumfrac} and \ref{fig:FeIIlumfrac-depl}, where we show the [Fe\ {\sc ii}] 1.644 $\umu$m line emission fraction as a function of pre-shock ionization and plasma $\beta$), for solar abundances (Fig.~\ref{fig:FeIIlumfrac-dens}) and for iron depleted by a factor of 100 below solar abundance (Fig.~\ref{fig:FeIIlumfrac-dens-depl}). As for variations with pre-shock plasma $\beta$, for the purposes of our study, there is relatively little effect on the fraction of total luminosity emitted in [Fe\ {\sc ii}] 1.644 $\umu$m; the fraction is $\sim 10^{-2}$ for solar abundances, and $\sim 10^{-4}$ for an iron depletion factor of 100.

The total emission from these partially-ionized slow shocks arises from a non-linear, time-dependent ionization state. This situation comes about via competition between collisional ionization caused by the post-shock temperature jump, and recombination that follows as the gas cools quickly. The ionization balance is generally far from equilibrium and is somewhat dependent on the initial ionization fractions, as the shocks never reach the hot equilibrium post-shock phase seen in faster shocks \citep[e.g.][]{Ae08}. This makes the initial ionization state important in determining the final shock emission spectra. The cooling arises primarily from the collisional excitation of lines (e.g.~H~$\alpha$, Ly~$\alpha$) and two-photon continuum emission of hydrogen, as well as a large number of forbidden and fine-structure transitions in metal species such as neutral oxygen or Fe\ {\sc ii}. As the pre-ionization state changes, the availability of electrons and the changing mean molecular weight (which alters the post-shock temperature), combined with the changing initial values for ion abundances and magnetic pressure, means that the integrated contribution of a particular line in a particular species (such as [Fe\ {\sc ii}] 1.644 $\umu$m) can vary both up and down as the mix of competing processes and species change. 

For example, the density reached when an important ion is most abundant may vary between models, depending on the temperature profile and the contribution of magnetic pressure support. Density `quenching' via collisional de-excitation can change the emission from some lines but not others, depending on their critical densities (cf.~\citealt{DS03}). Therefore, in any given model, the emission contribution from species competing with [Fe\ {\sc ii}] 1.644 $\umu$m emission will change, and so the Fe\ {\sc ii} cooling fraction will also change. Fortunately, the overall integrated variations in the [Fe\ {\sc ii}] fraction range over only factors of a few for the range of pre-shock conditions considered here. The relatively small variations in the initial CIE Fe\ {\sc ii} fractions in Table \ref{tab:CIE} may contribute to the rough stability of the [Fe\ {\sc ii}] 1.644 $\umu$m emission efficiency.

Additionally, the collisional excitation of neutral hydrogen in particular, enhancing H~$\alpha$, Ly~$\alpha$ and the two-photon continuum are in many models a dominant coolant, and as the neutral hydrogen species is common throughout all the models, the overall cooling is not as variable as it would be if the cooling were dominated by fleeting species. Likewise, the Fe\ {\sc ii} ion is often the most common Fe ion in the models. However, in any given model, the detailed efficiency outcome is difficult to predict in this highly non-linear system, so the variations seen in Figs.~\ref{fig:FeIIlumfrac}, \ref{fig:FeIIlumfrac-depl}, \ref{fig:FeIIlumfrac-dens}, and \ref{fig:FeIIlumfrac-dens-depl} are best interpreted in general terms, as specifics are difficult to isolate and prove.

\end{document}